\documentclass{aa}  
\usepackage[utf8]{inputenc}
\usepackage[T1]{fontenc}
\usepackage{amsmath,amssymb,amstext}
\usepackage{appendix}
\usepackage{babel}
\graphicspath{{img/}}
\usepackage{hyperref}
\hypersetup{colorlinks=true,citecolor=blue,linkcolor=blue!80!black,urlcolor=blue!80!black}
\usepackage{tabularx}
\usepackage{xcolor}
\usepackage{xspace}
\usepackage{threeparttable}
\usepackage{placeins}
\usepackage{dblfloatfix}    
\usepackage{graphicx}
\usepackage{subcaption}
\usepackage{txfonts}

\newcommand{\eq}[1]{Eq.\,(\ref{#1})}

\newcommand{\fig}[1]{Fig.\,\ref{#1}}

\newcommand{\figures}[2]{Figs.\,\ref{#1} to \ref{#2}}
\newcommand{\sect}[1]{Sect.\,\ref{#1}}
\newcommand{\app}[1]{Appendix\,\ref{#1}}
\newcommand{\tab}[1]{Table\,\ref{#1}}

\def\Rearth{R$_\oplus$\xspace}

\def\Rsun{R$_\odot$\xspace}
\def\Rc{\ensuremath{\mathrm{R}_\mathrm{c}}\xspace}
\def\Mc{\ensuremath{\mathrm{M}_\mathrm{c}}\xspace}
\def\Rp{\ensuremath{\mathrm{R}_\mathrm{p}}\xspace}
\def\Mp{\ensuremath{\mathrm{M}_\mathrm{p}}\xspace}
\def\Rstar{\ensuremath{\mathrm{R}_\mathrm{\ast}}\xspace}

\def\Tstar{\ensuremath{\mathrm{T}_\mathrm{\ast}}\xspace}
\def\Teq{\ensuremath{\mathrm{T}_\mathrm{eq}}\xspace}
\def\Tsurf{\ensuremath{\mathrm{T}_\mathrm{surf}}\xspace}
\def\Tdiff{\ensuremath{\mathrm{T}_\mathrm{diff}}\xspace}
\def\Pmineral{\ensuremath{\mathrm{P}_\mathrm{mineral}}\xspace}
\def\Pvap{\ensuremath{\mathrm{P}_\mathrm{vap}}}
\def\ph{\ensuremath{\mathrm{P}_\mathrm{H}^0}\xspace}
\def\cnc{55 Cnc e\xspace}
\def\atmo{\texttt{ATMO}\xspace}
\def\magmav{\texttt{MAGMAVOL}\xspace}
\def\vaporock{VapoRock\xspace}
\def\Sum{\textbf{Sum}\xspace}
\def\Equilibrium{\textbf{Equilibrium}\xspace}
\def\vaporj{g_j}
\def\liquidj{\ell_g}
\def\OT{\ensuremath{\mathrm{O_2}}\xspace}
\def\O{\ensuremath{\mathrm{O}}\xspace}
\def\H{\ensuremath{\mathrm{H}}\xspace}
\def\K{\ensuremath{\mathrm{K}}\xspace}
\def\Na{\ensuremath{\mathrm{Na}}\xspace}
\def\Mg{\ensuremath{\mathrm{Mg}}\xspace}
\def\Fe{\ensuremath{\mathrm{Fe}}\xspace}

\def\HT{\ensuremath{\mathrm{H_2}}\xspace}
\def\HTO{\ensuremath{\mathrm{H_2O}}\xspace}
\def\SiOT{\ensuremath{\mathrm{SiO_2}}\xspace}
\def\SiO{\ensuremath{\mathrm{SiO}}\xspace}
\def\SiHF{\ensuremath{\mathrm{SiH_4}}\xspace}
\def\SiHT{\ensuremath{\mathrm{SiH_3}}\xspace}
\def\SiHt{\ensuremath{\mathrm{SiH_2}}\xspace}

\begin{document}
\title{
Hydrogenated atmospheres of lava planets: atmospheric structure and emission spectra
}

\titlerunning{Hydrogenated atmospheres of lava planets}

\author{Aur\'{e}lien Falco\inst{1,2}
\and
Pascal Tremblin\inst{3}
\and
S\'{e}bastien Charnoz\inst{1}
\and Robert J. Ridgway \inst{4}
\and Pierre-Olivier Lagage \inst{2}
}

   \institute{(1) Universit\'e de Paris Cité, Institut de Physique du Globe de Paris, CNRS
F-75005 Paris, France\\
              \email{falco@ipgp.fr}\\
    (2) Laboratoire AIM, CEA, CNRS, Univ. Paris-Sud, UVSQ, Université Paris-Saclay, F-91191 Gif-sur-Yvette, France  \\
    (3) Maison de la simulation, CEA, CNRS, Univ. Paris-Sud, UVSQ, Université Paris-Saclay, F-91191 Gif-sur-Yvette, France  \\
    (4) Physics and Astronomy, Faculty of Environment, Science and Economy, University of Exeter, Exeter, EX4 4QL, UK \\
 }

\date{\today}

\abstract{
Ultra hot rocky super-Earths are thought to be sufficiently irradiated by their host star to melt their surface and thus allow for long-lasting magma oceans.
Some processes have been proposed for such planets to have retained primordial hydrogen captured during their formation while moving inward in the planetary system.
The new generation of space telescopes such as the James Webb Space Telescope may provide observations that are precise enough to characterize the atmospheres and perhaps the interiors of such exoplanets. 
We use a vaporization model that calculates the gas-liquid equilibrium between the atmosphere (including hydrogen) and the magma ocean, to compute the elemental composition of a variety of atmospheres with different quantities of hydrogen.
The elemental composition is then used in a steady-state atmospheric model (namely, ATMO), to compute the atmospheric structure and generate synthetic emission spectra.
With this method, we confirm previous results
showing that silicate atmospheres exhibit a thermal inversion, with most notably an emission peak of SiO at 9~$\mu m$.
We compare our method to the literature on the inclusion of hydrogen in the atmosphere, and show that hydrogen reduces the thermal inversion, because of the formation of $\HTO$, which has a strong greenhouse potential.
However, planets that are significantly irradiated by their host star are sufficiently hot to dissociate $\HTO$ and thus also maintain a thermal inversion.
The observational implications are twofold: 1) \HTO is more likely to be detected in colder atmospheres; 2) Detecting a thermal inversion in hotter atmospheres does not a priori exclude the presence of H (in its atomic form).
Due to the impact of H on the overall chemistry and atmospheric structure (and therefore, observations), 
we emphasize the importance of including volatiles in the calculation of the gas-liquid equilibrium.
Finally, we provide a criterion to determine potential targets for observation.
}

\maketitle

\section{Introduction}
\label{introduction}

The current developments in observational capabilities, including the launch of new telescopes such as the James Webb Space Telescope (JWST), has led to the study of relatively new categories of exoplanets that were not easily observable before.
More than just the detection of species in the atmosphere of exoplanets, observations may now enable a better characterization of the atmospheric structure and composition of these objects.
These include ultra hot rocky exoplanets (with equilibrium temperature > 1500~K).
This category of exoplanets have radii~<~5~\Rearth (Earth radii) and short periods (< 10 days) \citep{essack2020low}.
Their surface is strongly irradiated by their host star and can therefore be partially or fully melted.
They can be subdivided into sub-categories such as `lava worlds', `magma ocean worlds', or `highly volcanic planets', depending on the percentage of surface covered by lava/magma oceans \citep{Henning2018}.
In this study we primarily focus on super-Earths/sub-Neptunes that fit into the `lava worlds' category, i.e., we assume a fully melted surface, 
though we intend to study partially melted surfaces in the near future.

According to Fig.~11 from \cite{lebrun2013thermal}, the magma ocean phase would last longer than 100~Myr for planets that have orbital radii < 0.66~au. 
The possibility of a (partially or totally) melted surface offers an interesting setting in which the atmosphere is strongly influenced by the chemical composition of the magma ocean \citep{Dorn2021Hidden}. This allows for the inference of interior composition through the study of the atmosphere composition, obtained via the observations.
This could potentially also offer insight on early Earth characteristics, as magma oceans are speculated to be common in the past of large rocky planets \citep{schaefer2018magma,greenwood2005widespread}.

Although recent studies seem to report that temperate and rocky planets are not expected to have an observable atmosphere using current instrumentation \citep{Kreidberg_2019,zieba2022,Crossfield_2022,keles2022}, the high temperatures at the surface of these planets, due to their proximity with the star, should be enough to melt the surface into a magma ocean and create a thin silicate atmosphere \citep{Schaefer_2009,Miguel_2011,ito2015,Kite_2016,zilinskas2022}.
\cite{Schaefer_2009} found silicate atmospheres would be composed primarily of Na, O$_2$, O and SiO.
They also suggest that large Na and K clouds could surround these hot super-Earths.
\cnc and HD~149026~b should also allow the formation of mineral clouds in their atmosphere, according to \cite{mahapatra2017cloud}. 
At the same time, they also find that the high temperatures on some hot-rocky super-Earths (e.g. the day-side of Corot-7~b) result in an ionised atmospheric gas, and as they prevent
gas condensation, it is unlikely that clouds form on their day-side.
Studying pure Na, SiO and \SiOT atmospheres coupled with a magma ocean, \cite{nguyen2020} found that a steady-state pure Na atmosphere would not be sustainable and the surface would change over time.
\cite{zieba2022} suggest that only K2-141~b and \cnc are hot enough to have a molten dayside, among the four small ultra-short-period planets observed by the JWST during its Cycle 1 General Observers program (also including LHS~3844~b and GJ~367~b).
According to \cite{zilinskas2022},
the spectral energy distribution coming from
hotter stars is much more weighted towards shorter wavelengths,
and
the shape of
the T-P profile of lava worlds (with a silicate atmosphere) is determined by the shortwave/IR absorption of
stellar irradiation.

\cite{Otegi_2020} suggest that volatile-rich and rocky atmospheres can be separated using their density.
For example, they characterize \cnc as a rocky planet (density of 5.9~g/cm$^3$ \citep{Crida_2018_55CNCe}).
According to \cite{Rogers_2021}, planets under 2~\Rearth would be likely be stripped of $\HT$.
However, some rocky planets have an unusually low density, such as TOI-561~b (density of $\sim$4.4~g/cm$^-3$), which might be explained by an important volatile reservoir in their mantle \citep{piette2023rocky}.
Since the volatiles should escape to space quite rapidly, the presence of such volatiles should be linked to an important volatile inventory in the interior, which may have been captured during their formation.

Terrestrial planets may have migrated inward during their formation due to efficient disk-planet interaction, so highly irradiated magma ocean exoplanets could have formed further away at much lower temperatures and with much higher hydrogen budgets than they have today \citep{charnoz2023effect}.
Newly formed planets may gravitationally capture a  primordial atmosphere, rich in $\HT$, \HTO, and heavier elements, but largely dominated by mass by hydrogen \citep{Fegley_2020,Kite_2021}, though hydrogen should amount to no more than 6\% of the planet's mass through degassing \citep{Elkins_2008}.
For planets on temperate orbits, $\HT$ is expected to be a strong contributor to the duration of the magma ocean phase \citep{Lichtenberg_2021}, while \HTO, CO$_2$ and CH$_4$ are expected to have less effect, and
$\OT$, N$_2$ and CO are even less effective at keeping a magma ocean from solidifying.
\cite{hamano2013emergence} also show that increasing the \HTO content by tenfold would approximately increase the time before solidification by the same order of magnitude, due to the fact that the planet must lose water to cool down.
\cite{Dorn2021Hidden} suggest that even small amounts of greenhouse gases (few tens of bars of \HTO or $\HT$ for example) would melt the surface rocks of super-Earths and thus allow a magma ocean underneath the surface.
They hypothesize that water could still be present in the interior of \cnc, in which case they predict a non-zero partial pressure of water in a metal-rich atmosphere.
More precisely they forecast a 5\% mass fraction of \HTO, partitioned between a wet magma ocean and a steam atmosphere, which could explain the relatively low bulk density of the planet.
Finally, \cite{Kite_2020} use a Fe–Mg–Si–O–H model to study sub-Neptunes and the absorption of volatiles in the magma.
They found that on this type of planets, insulation due to the atmosphere would allow magma oceans to last indefinitely, and that the composition of the atmosphere is greatly influenced by the atmosphere-magma interaction.
According to \cite{Kite_2021}, the H$_2$ atmosphere lost is then replaced by a H$_2$O atmosphere, meaning the planet will not be a bare rock, and maintains a 150-200~km thick atmosphere.

Two effects can therefore allow for a magma ocean to last longer than the initial stage of planetary evolution: the stellar irradiation for ultra-short-period planets, and the presence of \HT or \HTO in the atmosphere.
In the current study, we are interested in the influence of hydrogen on the thermal structure of the atmosphere and the surface temperature of the planet, as well as its impact on spectral observations.

Larger bodies with a $\HT$ atmosphere could be easily probed via transmission spectroscopy \citep{Hu_2021}, but smaller bodies, or silicate atmospheres would require studies to be made using emission spectroscopy, characterizing the day-side of the planet, on which the atmosphere could be contained, when the planet is close to the secondary eclipse.
Observations with the JWST in this context will still remain a challenge \citep{zilinskas2023observability}.
We focus therefore in this study on emission spectroscopy of hot rocky worlds with or, without an hydrogenated atmosphere on top of a magma ocean and characterize their atmospheric structure and emission.



This study, which focuses on the atmospheric structure and observational implications of hot rocky super-earths, follows the method described in \cite{charnoz2023effect} for the computation of the gas-liquid equilibrium between a magma ocean and an atmosphere which contains a varying quantity of hydrogen, from H-poor to H-rich. 
As we focus on ideal gases, our study is only suited for planets with less than $10^5$ bars of H, so we restrict ourselves to rocky and sub-Neptune planets.
We will discuss in \sect{sec:vaporization_model} how the equilibrium is computed and how we take into account the effect of hydrogen,
which corresponds to the method we have described in \cite{charnoz2023effect}.
We compare this method (taking into account the gas equilibrium between the vapor and volatiles) to the sum of the vapor with volatiles, a method commonly used by the literature for example in \cite{zilinskas2023observability,piette2023rocky} in \sect{sec:sum_equilibrium}.
The atmospheric structure model will then be presented in \sect{sec:atmospheric_model}.
In \app{sec:silicate_case}, we compare our model to LavAtmos \citep{van2022lavatmos} for atmospheres completely deprived of hydrogen (silicate case).
Atmospheres and spectra for cases with hydrogen are discussed in \sect{sec:hydrogen_case}.
We propose a criterion for selection of potential candidates for observation in \sect{sec_hydrogen_index}.

\section{Gas-liquid equilibrium}
\label{sec:vaporization_model}

We have discussed in \cite{charnoz2023effect} the method used for the calculation of the elementary content of the vapor arising from the magma ocean, in presence of a pre-existing hydrogen content.
This method will be referred to as the \magmav (MAGMa+Atmospheric VOLatiles) method all along this study.
We focus on a Na-K-Mg-Al-Fe-Si-O+H system, where H is present only in the atmosphere and is in equilibrium with the vapor released by the magma ocean. 
We summarize the main aspects here: the atmosphere is assumed to  be at chemical equilibrium, so for a given elemental composition (number of moles of each atom) the molecules present for a given (P,T) are computed with the chemical equilibrium code CEA  \citep{Gordon_McBride_1996_CEA_code}.
In order to appropriately consider the interaction with the liquid magma ocean, partial pressures of atmospheric evaporated species (such as SiO, SiO$_2$, Mg, Na, K,Fe, O$_2$) must satisfy a number of vaporisation reactions (these reactions are listed in  \tab{table:Table_reaction_liquid_gas}) ensuring gas-liquid equilibrium. More specifically, all reactions, $j$ listed in Table \tab{table:Table_reaction_liquid_gas} imply an equilibrium reaction that reads :
\begin{equation}
    P_{\vaporj}=\frac{K_j(T)a(\liquidj)}{P_{\OT}^{s_j}},
    \label{eq_partial_pressures_equil}
\end{equation}
 where $\vaporj$ stands for the vapor species that bears the metal in the reaction $j$ (for example $Na_{(g)}$ in reaction $\#3$), $\liquidj$ is the corresponding liquid oxide ($NaO_{0.5}$ for reaction $\#3$), $s_j$ is the stoichiometric coefficient (1/4 in reaction $\#3$),
 \begin{table}[ht]
    \caption{Liquid-gas reactions}
    \label{table:Table_reaction_liquid_gas}
  \begin{threeparttable}
    \begin{tabular}{c|c|c|c} 
     \textbf{$\#$} & \textbf{Reaction } & \textbf{s} & \textbf{ref  }\\
       \hline
            1 & $\SiO_{2(\ell)} \Leftrightarrow \SiO_{(g)}+1/2 \OT$ &  1/2 & Chase(1998)    \\
            2 & $\SiO_{2(\ell)} \Leftrightarrow \SiO_{2(g)}  $ & 0 & Chase(1998)     \\
            3 & $\Na\O_{1/2(\ell)} \Leftrightarrow \Na_{(g)}+1/4 \OT$ & 1/4   & Chase(1998) \\
            4 & $\K\O_{1/2(\ell)} \Leftrightarrow \K_{(g)}+1/4 \OT$ & 1/4  & Chase(1998)   \\
            5 & $\Mg\O_{(\ell)} \Leftrightarrow \Mg_{(g)}+1/2 \OT$ & 1/2  & Chase(1998)     \\
            6 & $\Fe\O_{(\ell)} \Leftrightarrow \Fe_{(g)}+1/2 \OT$ & 1/2  & Chase(1998)     \\
        \end{tabular}
        \begin{tablenotes}\footnotesize
        \item[] On the left-hand side are the liquid oxides and on the right the gas oxides. Note that for species Na$_2$O and K$_2$O we have considered the oxide normalized by the number of metal atoms, as in \cite{Sossi_2020, sossi_etal2019}. The mole fraction of $\Na\O_{1/2}$ in melt is of $\sqrt{\Na_2\O}$. The same applies to K$_2$O. Values of all thermodynamic constants were taken from \cite{Chase-JANAF}. 
        \end{tablenotes}
  \end{threeparttable}
\end{table}
$K_j(T)$ is the reaction coefficient, $a(\liquidj)$ is the activity of liquid $\#j$, that is $a(\liquidj)=X_j\times\Gamma_j$ where $X_j$ is the mole fraction of $\liquidj$ in the magma ocean (assumed to be fixed) and $\Gamma_j$ is the activity coefficient of $j$ in the liquid due to non ideal mixing effects. Activity coefficients, $\Gamma_j$, that are specific to the composition of the liquid, are interpolated using outputs of the \vaporock code \citep{wolf2022vaporock},
our procedure is explained in details in \cite{charnoz2023effect}.  
Initially, the atomic molar fractions and total pressure of the atmosphere are unknown variables. 
Their value is searched with an iterative procedure to satisfy all the above equations.

The oxygen fugacity (f$_\OT$) is computed by assuming congruent evaporation of melt oxides, so that f$_\OT$ is found by solving for mass conservation of oxygen as in \cite{charnoz2023effect} or in \cite{van2022lavatmos}. 
This is justified because our composition of planetary melted mantle is BSE, where all iron is under a single form, FeO, so there is no pressure buffer. 
The effective f$_\OT$ can be found by checking the \OT partial pressure displayed in
Fig~B.1~to~B.4 in \cite{charnoz2023effect}.
More details can be found in \cite{charnoz2023effect}. 
It's important to highlight that our code does not consider the dissolution of volatiles in the liquid, such as \HTO. Instead, it computes the equilibrium between the magma ocean (with a fixed composition containing only silicate oxides) and an atmosphere, wherein we specify a particular hydrogen content \ph.

Fig~B.1~to~B.4 of \cite{charnoz2023effect} show 
the partial pressures of most abundant species in the vapor at the surface above the magma ocean, when the atmosphere is in equilibrium with the magma ocean.
They are displayed for different values of the monoatomic pressure of H, \ph (i.e., the pressure of hydrogen if all atoms of hydrogen were under the form of H).
We call \Pmineral the sum of all partial pressures for an atmosphere devoid of H and containing only evaporated species. 
At low H content, $P_H^0 < 0.2$~bar, the atmosphere is not affected by the presence of H, and we have a close-to pure mineral atmosphere, with an almost constant total pressure $\Pvap\sim \Pmineral$.
At $P_H^0 > 0.2$~bar, the atmosphere becomes strongly hydrogenated and the atmospheric composition drastically changes. 
\Pmineral depends on the temperature and for a planet with Bulk Silicate Earth composition, we have found that \Pmineral (expressed in bar) is well fitted by:
\begin{equation}
\log_{10}(\Pmineral(T))=-4.02\times 10^{-7} T^2+  5.26\times 10^{-3} T -12.75, 
\label{eq_pmineral}
\end{equation}
with T in Kelvins, using data from \cite{charnoz2023effect}.

At low hydrogen budgets, i.e., \ph~<~0.1~bar, the dominant species are SiO and Na, closely followed by \OT. 
Fe is also quite abundant.
For a higher hydrogen budget, i.e., 1~<~\ph~<~$10^3$~bar, \HT starts to dominate, followed by \HTO, SiO and H, which become relatively less abundant as the partial pressure of \HT increases.
Na declines notably faster than other species. 
For a very high hydrogen budget, i.e., \ph~>~$10^4$~bar, species like \SiHF, \SiHT and \SiHt start to appear, and even become quite abundant.
Especially \SiHF is the third most abundant species after \HT and \HTO for very high \ph.

\begin{figure}[t]\centering
\begin{minipage}{.37\textwidth}
      \includegraphics[width=\textwidth]{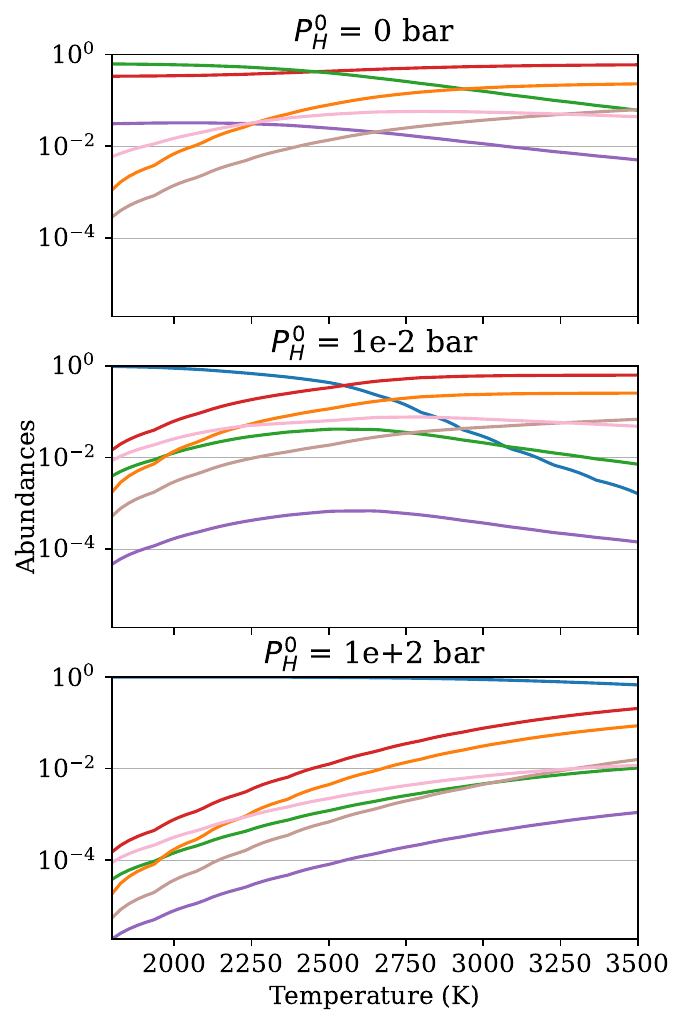}
\end{minipage}
\begin{minipage}{.1\textwidth}    
\includegraphics[width=.8\textwidth,keepaspectratio]{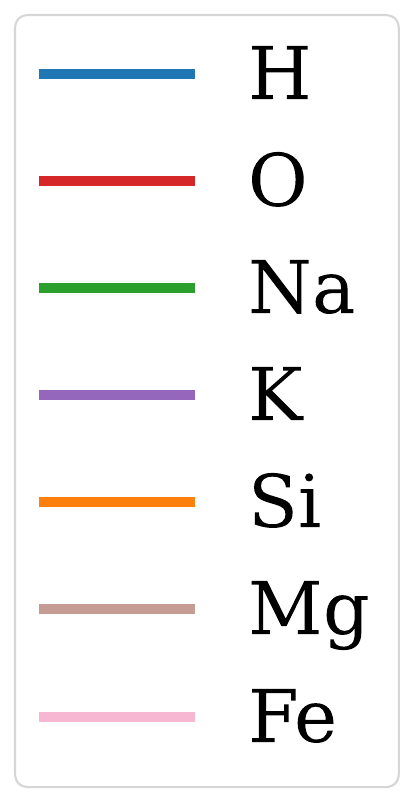}
\end{minipage}
  \vspace{-1mm}
  \caption{Vapor element abundances, extracted from the output of \magmav (see \cite{charnoz2023effect}).
  The magma ocean releases more and more refractory species (Si, O, Mg, ...) with increasing temperatures. The relative abundance of H thus decreases.
  }
  \label{fig:element_abundances}%
\end{figure}

The molar fractions of all considered atoms  are shown in \fig{fig:element_abundances} for different \ph. As we consider the atmosphere is well mixed vertically, the same atomic molar fractions will be used at every altitude in our vertical structure code.
Note that hydrogen is not present when \ph is equal to 0.
For \ph~=~0 (top plot), the most abundant element is Na for colder temperatures, but O becomes dominant for hotter temperatures (around 2500-3000~K), closely followed by Si.
K follows the same trend as Na but is less abundant, while Mg and Fe follow the trend of Si but also less abundant.
For \ph~=~$10^{-2}$~bar, the dominant species at low temperatures is H, followed by O.
The relative abundance of H is decreasing with increasing temperatures as heavy atoms (Si, Mg, Na, K) are released by the magma ocean.
The more volatile species such as Na, K, and Fe follow the same trend for temperatures > 2500~K.
The dominant species for T~>~2500~K becomes O, followed by Si.
For \ph~=~$10^2$~bar, the vapor is dominated by H, the second and third most abundant species are O and Si, respectively, which become more and more preponderant in the mix with increasing temperatures.

\section{Radiative transfer \& atmospheric structure}
\label{sec:atmospheric_model}

The atmospheric structure is computed via \atmo \citep{Amundsen2014,Amundsen2017,Tremblin_2015_ATMO,tremblin2016cloudless,Drummond2016} using the element abundances computed by the vaporization code (\sect{sec:vaporization_model}).
\atmo
solves for the pressure-temperature structure of an atmosphere by finding the energy flux balance in each model level, i.e.,
\begin{equation}
\label{equation:energy}
  \int_0^\infty\left(F_\mathrm{rad}(\nu) +F^0_\mathrm{star}(\nu)e^{\tau_\nu/\mu_\mathrm{star}}\right)d\nu + F_\mathrm{conv} = \sigma T_\mathrm{int}^4,
\end{equation}
where $F_\mathrm{rad}(\nu)$ and
$F_\mathrm{conv}$ are the spectral radiative flux and the convective flux,
respectively, and $\tau_\nu$ is the vertical monochromatic optical depth. $T_\mathrm{int}$ is the internal temperature of the object
corresponding to the surface flux at which the object cools in the
absence of irradiation; $T_\mathrm{int}$ is equivalent to the effective temperature $T_\mathrm{eff}$ in the absence of irradiation. $F^0_\mathrm{star}(\nu)$ is the incoming
irradiation flux from the star at the top of the atmosphere and
$\mu_\mathrm{star}$ = $\cos\theta$ where $\theta$ is the angle of incoming radiation off the vertical.

As mentioned before, we focus on emission spectroscopy in this study.
We rely on correlated-k opacities, which are more accurate than cross-sections \citep{Amundsen2014}.
The list of opacities used is:
Na,
K,
Fe,
FeH,
\HT-\HT,
\HTO, 
MgO,
SiO,
\SiOT,
\SiHF,
\SiHt.
Optical absorbers include Na, K, Fe, SiO and MgO while \HTO will be the main source of infrared absorption.
The spectral range goes from 0.2 to 2000~$\mu$m, with a resolution of 5000 for the synthetic spectra while it is 32 for the computation of the PT profiles.
For details about the opacities, see \app{sec:opacity_sources}.

The chemistry of the upper layers of the atmosphere is assumed to be in equilibrium and is computed via \atmo.
The model does not include photochemical processes (this could be investigated in future studies).
The code has been adapted to take into account the vapor elemental composition calculated by the gas-liquid equilibrium model (\sect{sec:vaporization_model}).
This includes inserting in the model the vapor pressure \Pvap(T) computed by the model, i.e., the total sum of the partial pressures of all outgassed species.
The saturating-vapor pressure defines the limit between the ocean and the atmosphere, i.e when P is equal to the saturating vapor pressure of the magma ocean, we assume the atmosphere is in contact with the liquid and that this point defines the atmospheric base.
\begin{figure}[ht]\centering
  \includegraphics[width=.44\textwidth]{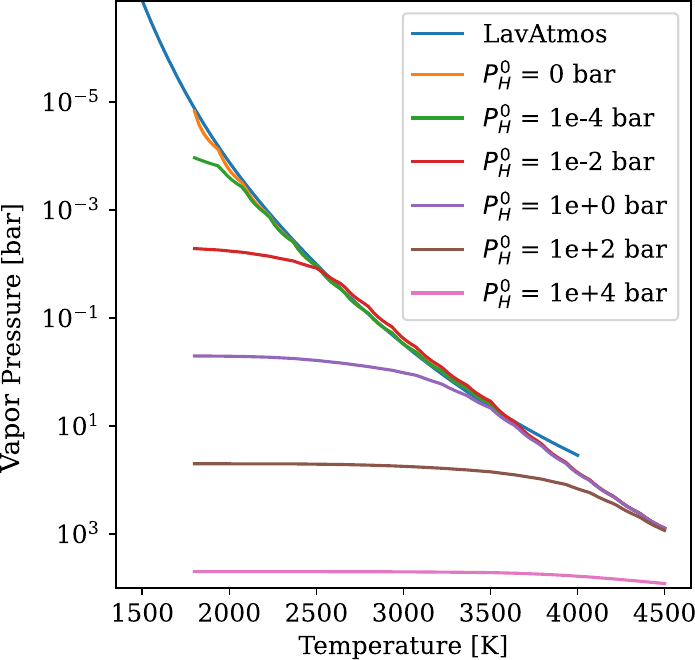}
  \caption{Vapor pressure \Pvap(T) computed by \magmav 
  for different temperatures T, and for different content values of hydrogen~\ph, with LavAtmos taken as a reference.}
  \label{fig:p_vap}
\end{figure}
It is a result of the \magmav code and is shown for different values of \ph in \fig{fig:p_vap}.
The more we add hydrogen, the higher the saturating vapor pressure is.
We have taken the LavAtmos code \citep{van2022lavatmos} as a reference for the hydrogen-free case.
LavAtmos is an open-source code that computes the chemical gas-liquid equilibrium, i.e., that is also able to compute the vapor content above a magma ocean.
A more extensive comparison of the \magmav and LavAtmos codes is discussed in \app{sec:silicate_case}
(along with a comparison of with MAGMA \citep{Schaefer_2009}). 
Both the LavAtmos and the \magmav codes show very similar trends.
We restrict this comparison to the hydrogen-free case as LavAtmos does not yet compute the equilibrium between the atmospheric volatiles, like H,  and the silicated vapor.

The atmospheric molecular composition is computed as follows:
for each iteration of the solver, we use the pressure-temperature profile to identify the point (P,T) where the profile intersects with the vapor pressure curve, i.e., we have P~=~\Pvap(T). 
We extract the element abundances at this temperature from the \magmav code (\fig{fig:element_abundances}).
The element abundances are injected into \atmo, which recomputes the chemistry of the upper layers.
For pressures higher than \Pvap(T), the medium is considered to be liquid in \atmo.
The ocean is assumed to be optically thick with a grey opacity of 1000~cm$^2$/g and zero albedo. We have explored the sensitivity of the model to this opacity value and it makes no noticeable difference on the PT structure once the opacity is high enough to absorb the radiation reaching the surface in the first layer of the ocean. We have assumed a zero convective flux in the ocean in this study: we tested this assumption by using a convective transport with an adiabatic index close to~1, it also makes no difference to the resulting PT structure.

\section{The impact of hydrogen on the atmospheric structure}
\label{sec:hydrogen_case}

We study here how the presence of hydrogen impacts the atmospheric structure, the surface temperature and the associated spectral observations.

\subsection{Study parameters}
\label{study_cases}
We focus on a planet with a radius equal to 2~\Rearth (Earth radii).
The stellar spectrum used corresponds to 55 Cnc.
The semi major-axis (or distance between the planet and the star) $D$ is defined using the equilibrium temperature \Teq of the planet, using:
\begin{equation}
    T_\mathrm{eq}^4 = (1 - A_p) f \frac{R_\ast^2}{D^2} T_\ast^4,
    \label{eq:teq}
\end{equation}
where \Tstar and \Rstar are the temperature and radius of the star, respectively, $f$ is the heat redistribution factor.
For this definition, we have used a uniform heat redistribution, f~=~1/4 (e.g. \cite{essack2020low}), and the albedo $A_p$ is considered to be zero.
\tab{tab:teq_d} can be used to quickly translate an equilibrium temperature from our results to an orbital distance.
\begin{table}[ht]\centering
    \caption{Equilibrium temperature \Teq and the corresponding distance $D$ to star, for \Tstar~=~5200~K and \Rstar~=~1.118~\Rsun (solar radii).
    }
    \label{tab:teq_d}
    \begin{tabular}{cc}
    \Teq (K) & $D$ (au) \\\hline
    2000~K & 0.0176 \\
    2400~K & 0.0122 \\
    2800~K & 0.009 \\
    \hline
    \end{tabular}
\end{table}
Since we have set the stellar parameters (to that of 55 Cnc), the higher \Teq corresponds to a smaller semi-major axis $D$.

\subsection{Atmospheric structure}

\begin{figure}[ht]\centering
  \includegraphics[width=.5\textwidth]{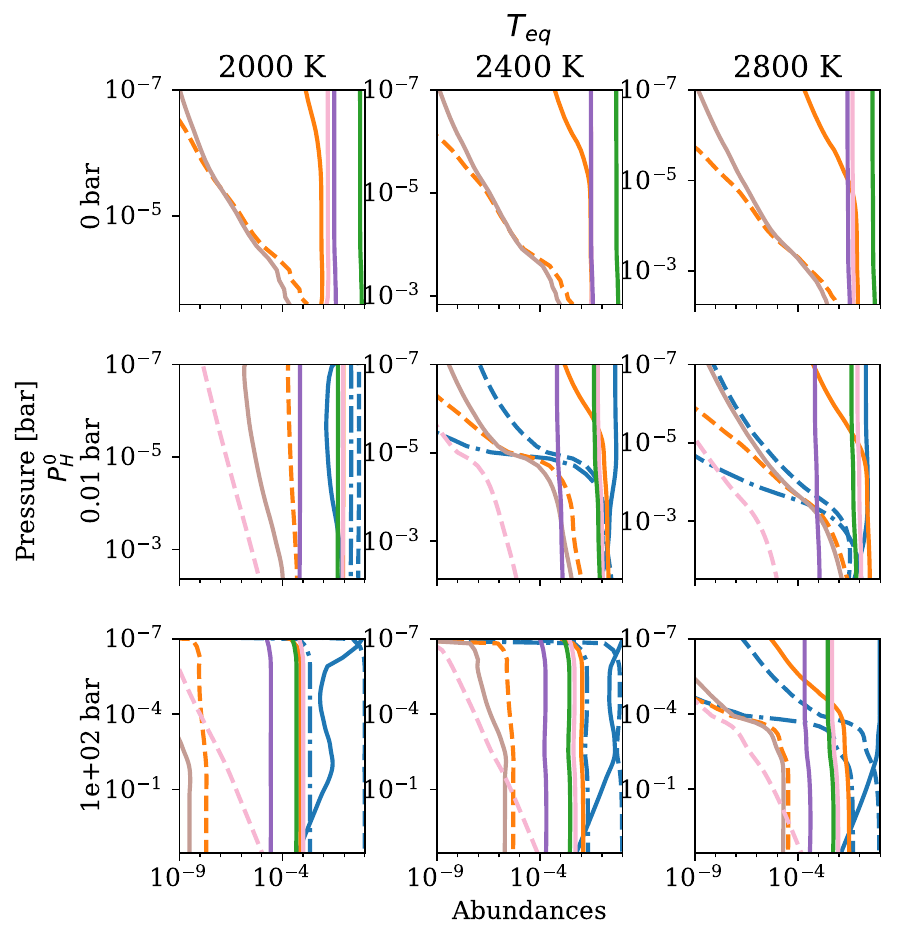}

  \includegraphics[width=.27\textwidth]{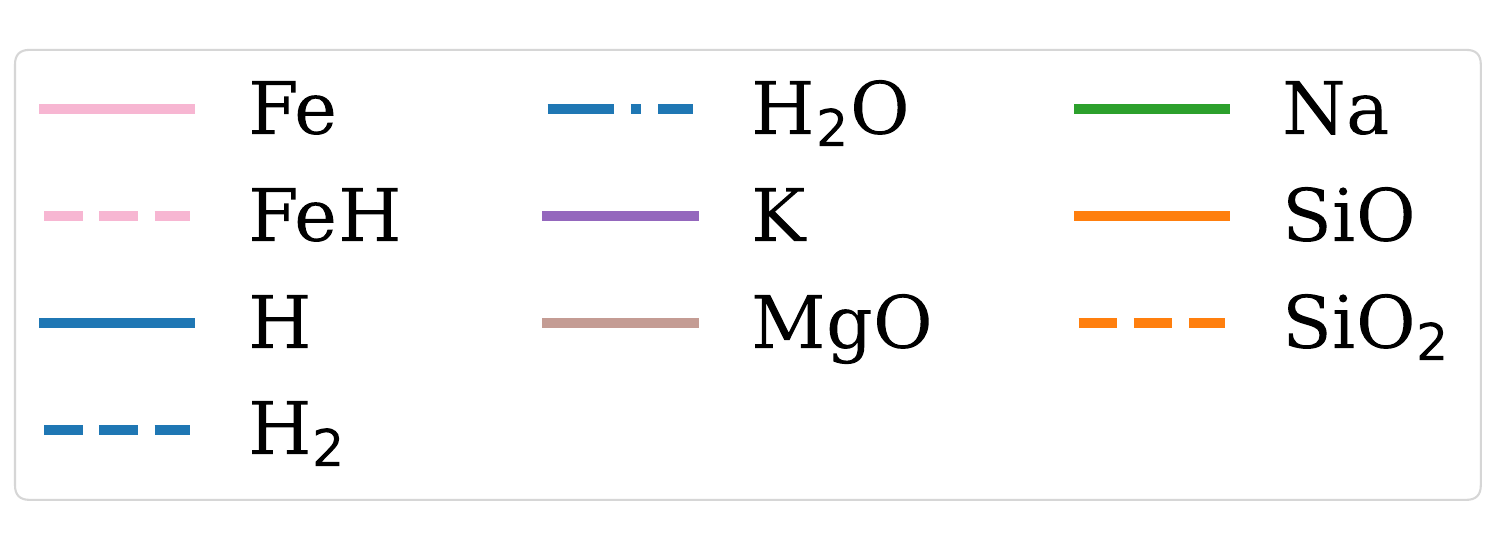}
  \vspace*{-2mm}
  \caption{Abundances in the atmosphere for cases with \ph~=0, ~$10^{-2}$ and $10^{2}$ bar (rows) and equilibrium temperatures of 
  $2000$~K,
  $2400$~K
  and $2800$~K (columns).
  The y-axis scale is different from one plot to the other; the minimum pressure is constant ($10^{-7}$~bar) while the surface pressure changes for each case.
  }
  \label{Fig_x}%
\end{figure}

\begin{figure}[ht]\centering
  \includegraphics[width=.4\textwidth]{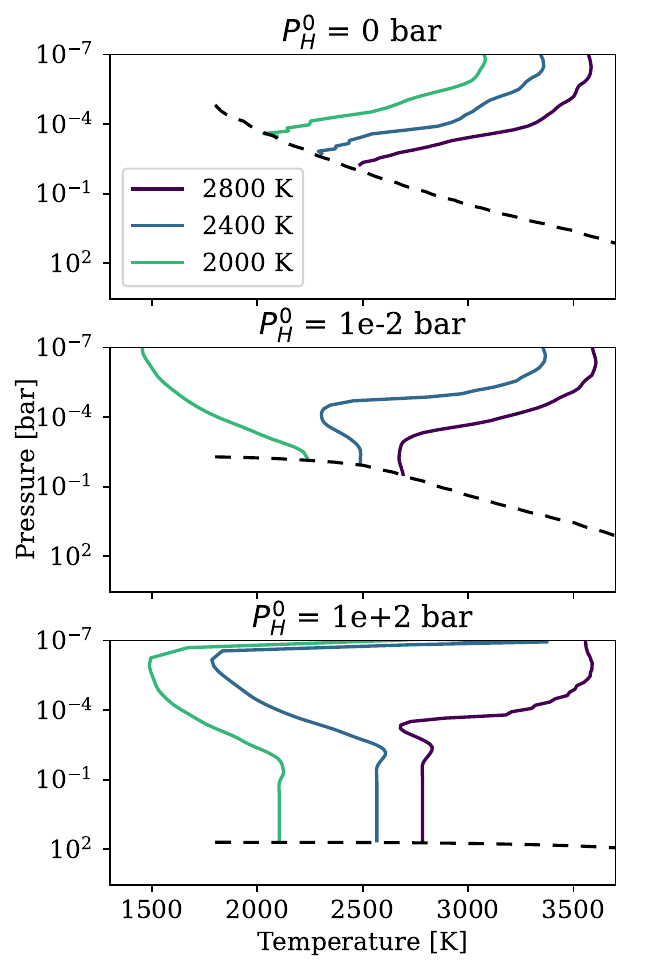}
  \caption{Thermal structure of the atmosphere calculated by \atmo (colored lines), for different monoatomic pressure of hydrogen \ph.
  The colors indicate the equilibrium temperature \Teq chosen for the simulation.
  The dotted lines show the vapor pressure corresponding to the limit between the magma ocean and the gaseous atmosphere, for each case.
  The silicate case (top plot) is more extensively discussed in \app{sec:silicate_case}. 
  }
  \label{Fig_PTs_phs}%
\end{figure}

The molecular composition calculated by \magmav corresponds to the gas-liquid interface.
However, it is not representative of the upper atmosphere as P and T change with altitude.
The atmospheric structure is given in \fig{Fig_x} (chemical composition) and \fig{Fig_PTs_phs} (PT profile).
The correlated-k opacities used in the atmospheric code are detailed in \sect{sec:opacity_sources} and have a spectral resolution of 32 points.

Examples are given here for cases with different hydrogen contents (\fig{Fig_x}). 
\app{sec:silicate_case} also shows a comparison to LavAtmos in the hydrogen-free case.
For cases with a very high hydrogen budget (\ph~>~$10^4$~bar), we will also discuss the abundance and the spectral features of \SiHF in \sect{sec:sih4}.

For lower equilibrium temperatures (see \eq{eq:teq}),
i.e., $\Teq~=~2000$~K in \fig{Fig_x}, 
the majority of the atmosphere is dominated by $\HT$
when $P_H^0 > 10^{-2}$~bar.
The second more abundant molecule is then \HTO at $P_H^0 = 10^{-2}$~bar.
SiO is slightly more abundant when $P_H^0 = 10^{-2}$~bar than in the pure silicate case, while
Na and K are much less abundant (while Na is by far the most dominant species in the silicate case).
The chemistry at such equilibrium temperatures (<~2200~K) is mainly driven by the evaporation of the magma ocean.
As can be seen in Fig.~8 of \cite{charnoz2023effect}, Na and K are less and less abundant when H increases. The abundance of SiO peaks at $P_H^0 = 10^{-2}$~bar and then decreases though at a lesser rate than Na and K.

The intermediate case (\Teq$ = 2400$~K) is very interesting, as it shows how the chemistry can drastically change when the hydrogen budget varies.
MgO, \HTO and $\SiO_2$ all start to thermally dissociate around $10^{-5}$~bar for \ph~=~$10^{-2}$~bar, while SiO dissociates only around $10^{-6}$~bar.
Note that we will refer here to \emph{dissociation} as the process of thermal dissociation of the
molecular species in the high atmosphere as a result of chemical
equilibrium in high temperature/low pressure environments.
We emphasize that there are no photochemical processes here and the chemistry is considered to be at equilibrium.
The chemistry of this range of equilibrium temperatures (>~2200~K) is now partially driven by the evaporation of the magma ocean but also by thermal dissociation.

For very hot planets (\Teq$ = 2800$~K), 
hydrogen-bearing species are more abundant in the deep atmosphere and dissociate into H in the upper atmosphere, the transition occurring between $10^{-3}$ and $10^{-5}$ bar.
SiO is abundant for pressures $> 10^{-5}$ bar and decreases in upper layers.
In the case of $P_H^0 = 10^{2}$~bar, \HTO is quite abundant (near $10\%$ of the mix), but dissociates also below $10^{-4}$~bar.

Chemistry and pressure-temperature profiles are inter-dependent.
\fig{Fig_PTs_phs} shows the pressure-temperature profiles of the atmosphere, for cases with different H content \ph, and different equilibrium temperatures \Teq. 
The dashed line shows the vapor pressure line \Pvap(T), the limit between the atmosphere and the magma ocean (extracted from \fig{fig:p_vap}).
As indicated in \sect{sec:atmospheric_model}, the region below the dashed line is treated as an ocean in the atmospheric model (\atmo), and the element abundances are derived from \fig{fig:element_abundances} at the temperature T corresponding to the intersection of the PT profile with \Pvap(T).

For hydrogen-free cases (top plot), the atmospheric structure exhibits a pronounced thermal inversion, i.e., the temperature increases with altitude. 
This is because the atmosphere is dominated by species that absorb in the visible (SiO, Na, K, Fe, ...).
This has been shown previously by \cite{ito2015,zilinskas2022}.
We refer the interested reader to \app{sec:silicate_case} and the above papers for more information.
Inverted PT profiles (which follow a thermal inversion) are linked to spectral emission features, 
while non-inverted PT profiles 
(for which the temperature decreases with altitude)
are linked to absorption features. 
This is of interest for future observations, as the difference between an inverted and a non-inverted profile is more easily detectable than to observe molecular spectral features.

\HTO being an infrared absorber has a greenhouse effect (reverse to the thermal inversion) when it is sufficiently abundant.
Therefore, cases with a higher hydrogen content \ph are less prone to exhibit a thermal inversion.
For example at \Teq~=~2000~K and \ph~>~$10^{-2}$~bar, the thermal inversion is completely removed, due to the fact that the atmosphere is dominated by $\HTO$ (see \fig{Fig_x}).

In instances where \Teq~=~2800~K, the thermal inversion persists for any hydrogen content.
Indeed, for $P_H^0 = 10^{-2}$~bar, \HTO dissociates 
at around $10^{-3}$~bar, and is therefore abundant enough only to reverse the thermal inversion in a very narrow pressure band between $10^{-1}$ and $10^{-3}$~bar.
At $P_H^0 = 10^{2}$~bar, \HTO dissociates at around $10^{-4}$~bar and is sufficiently abundant to eliminate the thermal inversion for pressures beyond $10^{-4}$~bar.
The thermal inversion materializes at lower pressures, when \HTO is dissociated.

\begin{figure}[ht]\centering
  \includegraphics[width=.4\textwidth]{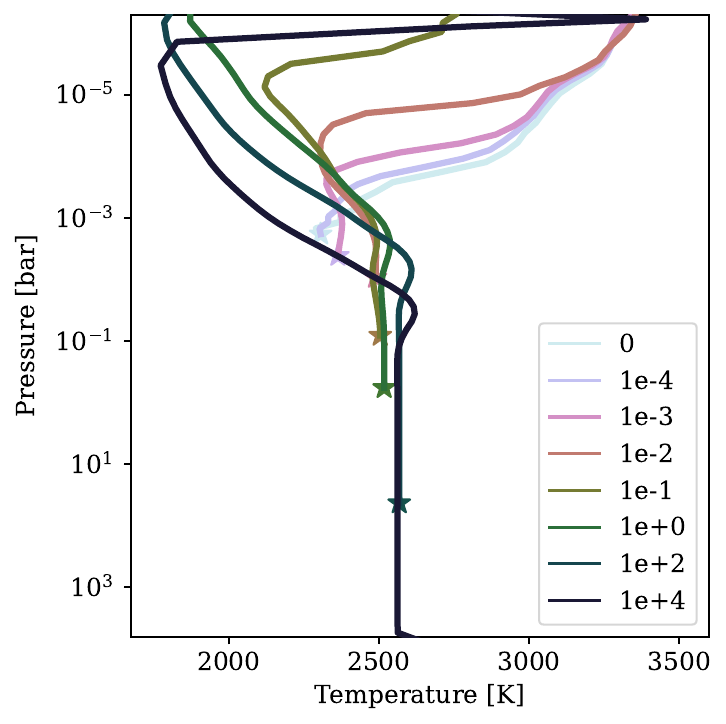}
  \caption{Thermal structure of the atmosphere of a rocky planet at \Teq = 2400~K  calculated by \atmo (plain lines), for different hydrogen content \ph (in bar), indicated by the colors.
  The stars indicate the surface of each case, i.e., the point where the atmosphere reaches the vapor pressure (see \fig{fig:p_vap}), and therefore, the ocean below.
  The surface temperature and pressure increase for larger \ph.
  }
  \label{Fig_pt_2400}%
\end{figure}

The case at \Teq~=~2400~K is very interesting as it seems to be at a critical point between the non-inverted and the thermal inversion regimes.
\fig{Fig_pt_2400} 
shows the pressure-temperature profiles at this equilibrium temperature, for the different values of hydrogen content \ph.
The color indicates the hydrogen content \ph of the atmosphere.
The surface is indicated by a star.
As the hydrogen content increases, we can see that the thermal inversion occurs at higher and higher altitudes until it disappears, replaced by an non-inverted profile.
This is closely tied to the presence of \HTO. 
In pressure ranges where \HTO dissociates, the thermal inversion regime sets in.
For the intermediate case with \ph~=~$10^{-2}$~bar, we can see on \fig{Fig_x} that
\HTO starts to dissociate around $10^{-5}$~bar, while SiO dissociates only around $10^{-6}$~bar.
The non-inverted profile occurs when \HTO
is abundant enough to be opaque,
between $10^{-3}$ to $10^{-5}$ bar, and above layers maintain a thermal inversion (when \HTO is dissociated).

\subsection{Thermal inversion}

We here simplify the definition of the thermal inversion and assume it is given by the difference between the surface temperature and the temperature at high altitude, 
where the atmosphere is not sufficiently opaque for radiative effects to be detectable:

\begin{equation}
    \Tdiff~=~\textnormal{T}(\tau_m=10^{-3}) - \Tsurf,
    \label{eq:tdiff}
\end{equation}
where $\tau_m$ is the mean optical depth for wavelengths between 0.5 and 20~$\mu m$.
Here we have chosen to set $\tau_m$ at $10^{-3}$.
The atmosphere becomes partially opaque in lower layers and starts to absorb incoming stellar flux.
As shown in \app{optical_depth}, $\tau_m=10^{-3}$ corresponds to pressures around $10^{-6}$ or $10^{-5}$~bar.
The surface pressure varies from case to case.

\begin{figure}[ht]\centering
  \includegraphics[width=.49\textwidth]{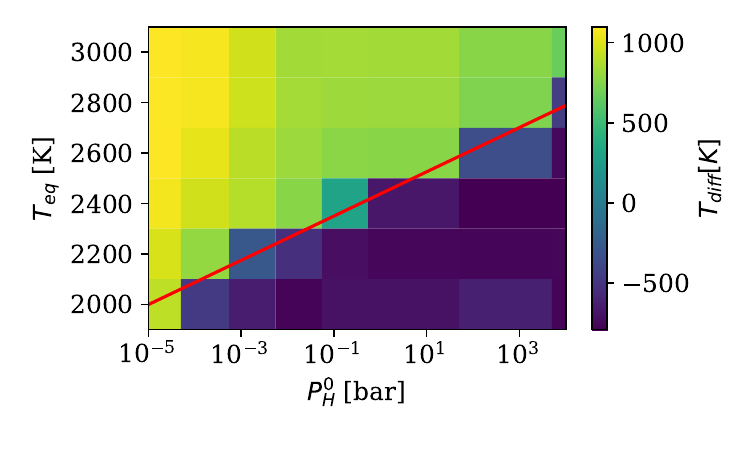}
  \vspace*{-1cm}
  \caption{Difference of temperatures \Tdiff (in color, defined in \eq{eq:tdiff}) at a pressure where $\tau_m=10^{-3}$ (around $10^{-6}$~bar) and the surface, for different 
  hydrogen content \ph and equilibrium temperature \Teq.
  Positive values indicate inverted PT profiles (presence of thermal inversion). 
  Negative values  indicate non-inverted PT profiles (absence of thermal inversion).
  The shift between the two regimes does not occur at the same hydrogen content for different equilibrium temperatures
  (hotter planets require more hydrogen to fall into the non-inverted category).
  This is indicated by the (fitted) red line, which has for formula \eq{eq_tdiff}.
  }
  \label{Fig_thermal_inversion}%
\end{figure}

We analyse the presence of a thermal inversion in the various cases proposed here, i.e., when varying the equilibrium temperature of the planet \Teq and its hydrogen content \ph.
Hotter planets can sustain a thermal inversion better than colder planets in the presence of hydrogen, as shown in \fig{Fig_thermal_inversion}.

The cases for which there is a thermal inversion either require a low content of hydrogen \emph{or} for the planet to be strongly irradiated (i.e., a large \Teq).
The thermal inversion is strongest (\Tdiff >~1000~K) in the cases with no hydrogen and a high equilibrium temperature.
The profile is non-inverted in the case with a lot of hydrogen and less irradiated (\Tdiff~$\sim$~-700~K). 
This effect is not linear.
The transition between a thermal inversion and a non-inverted profile is very sharp.
We fitted the transition by the red line in \fig{Fig_thermal_inversion}. Its formula is given by:
\begin{equation}
\Teq = 87.72*\log_{10}(\ph) + 2437.93
\label{eq_tdiff}
\end{equation}
This very sharp transition is linked to the strong thermal difference between the bottom of the atmosphere and the upper atmosphere. 
As more hydrogen is introduced, the thermal inversion occurs at progressively higher altitudes. 
When the thermal inversion occurs in non-opaque pressure ranges
(defined here as $\tau~<~10^{-3}$), 
the profile is classified as non-inverted.
Since the radiative effects are less strong for lower pressures (see \fig{Fig_optical_depth}), 
the main contribution to the emission will come from the zone between the surface and
the pressure level where $\tau~>~10^{-3}$ (P~>$\sim10^{-6}$~bar),
meaning if there is a thermal inversion occurring at lower pressures, it will not be linked to emission features (see \sect{sec_spectral_features}).

As shown by \fig{Fig_thermal_inversion} and \eq{eq_tdiff},
the hydrogen threshold between the two regimes (thermal inversion and non-inverted profile) depends on the equilibrium temperature.
For hotter planets, the hydrogen threshold is larger.
In our simulations, it occurs at a very low \ph for \Teq~=~2000~K, around $10^{-3}$~bar for \Teq~=~2200~K, 1~bar for \Teq~=~2400~K and 10~bar for \Teq~=~2600~K.

The fit of \eq{eq_tdiff} and the above values are to be taken with caution, as they will change for different mantle and volatile compositions (see also \sect{sec:discussion}).

\subsection{Temperature at the surface of the magma ocean}

\begin{figure}[ht]\centering
  \includegraphics[width=.48\textwidth]{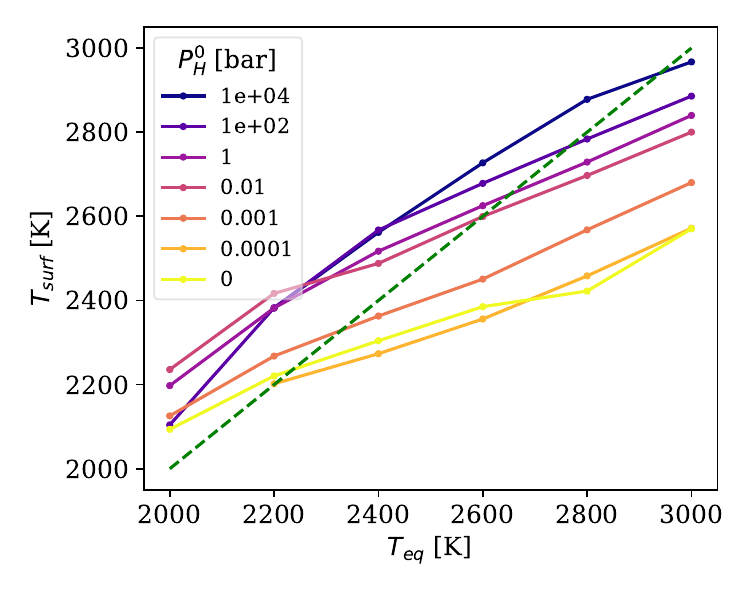}
  \vspace*{-2mm}
  \caption{Surface temperature \Tsurf for different equilibrium temperatures \Teq, depending on the hydrogen content of the atmosphere (\ph, in colors).
  The surface temperature may be underestimated for large \ph, as explained in \sect{sec:discussion} (see \cite{tremblin2017advection}).
  The green dashed line shows when \Tsurf = \Teq, for reference.
  }
  \label{Fig_tsurf}%
\end{figure}

The surface temperature is affected by the greenhouse effect of \HTO, and therefore increases with the amount of hydrogen in the system, as shown by \fig{Fig_tsurf}.
The difference between the surface temperature in the simulation with no hydrogen and the surface temperature in the simulation with a substantial amount of hydrogen for \Teq~=~2400~K is approximately 500~K for example.

This figure also shows a non-linearity in the increase of the surface temperature: 
there is a jump between \ph = $10^{-3}$ and $10^{-2}$ bar.
The measurement of hydrogen is therefore quite critical to properly evaluate the surface temperature.

For lower \Teq (<~2300~K), we can also see that there seems to be a limit to the increase of the surface temperature, which even decreases for very large hydrogen contents.
This is due to the outgassing of the magma ocean.
At \Teq=~=2000~K for example, \HTO is much less abundant when adding \ph~=~10$^2$~bar than at \ph~=~10$^0$~bar (see \fig{Fig_x}). 
This is better seen in Fig.~8 from \cite{charnoz2023effect}.
Using \fig{fig:element_abundances}, we can also see that at \ph~=~10$^2$~bar and 2000~K, O is much less abundant, which is the reason why H is primarily under the form of \HT rather than \HTO.
Due to its lower abundance, the presence of \HTO results in a milder greenhouse effect, leading to a less elevated surface temperature.
At 2400~K, \HTO is also less abundant, but the difference is much less ($x_\HTO > 10^{-2}$ at \ph~=~10$^2$~bar), which makes the greenhouse effect more effective and thus does increase the surface temperature.

In these simulations, an isothermal layer is present in the deep atmosphere for cases with a high content of hydrogen.
This could be an artefact from the 1D formulation, as will be discussed in \sect{sec:discussion} (see \cite{tremblin2017advection}).
The surface temperatures shown here may therefore be underestimated.

\subsection{Spectral features}
\label{sec_spectral_features}

\begin{figure}[ht]\centering
  \includegraphics[width=.5\textwidth]{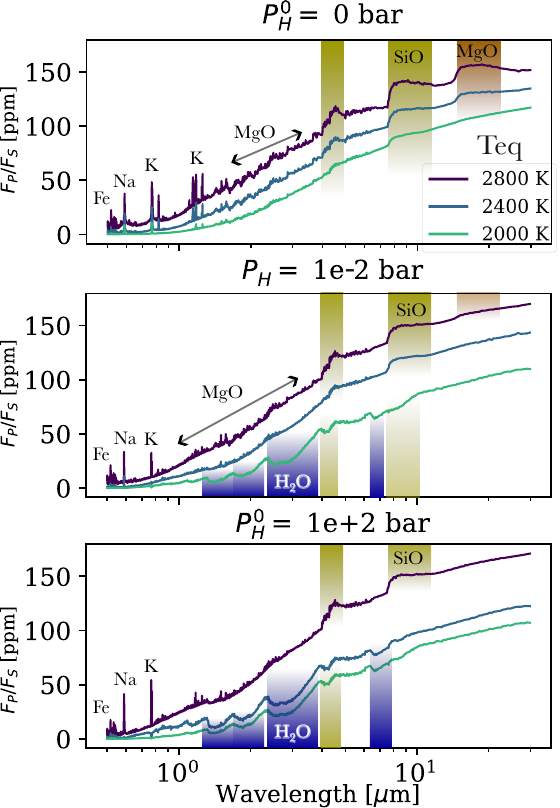}
  \caption{Emission spectra (planetary to stellar flux) for different equilibrium temperatures and compositions.
Emission features are highlighted by top-down color gradients, while absorption features are highlighted by bottom-up gradients. SiO is shown in yellow-ish color, MgO in brown-ish, and \HTO in blue.
}
  \label{Fig_spectra_phs}%
\end{figure}

\begin{figure}[ht]\centering
  \includegraphics[width=.5\textwidth]{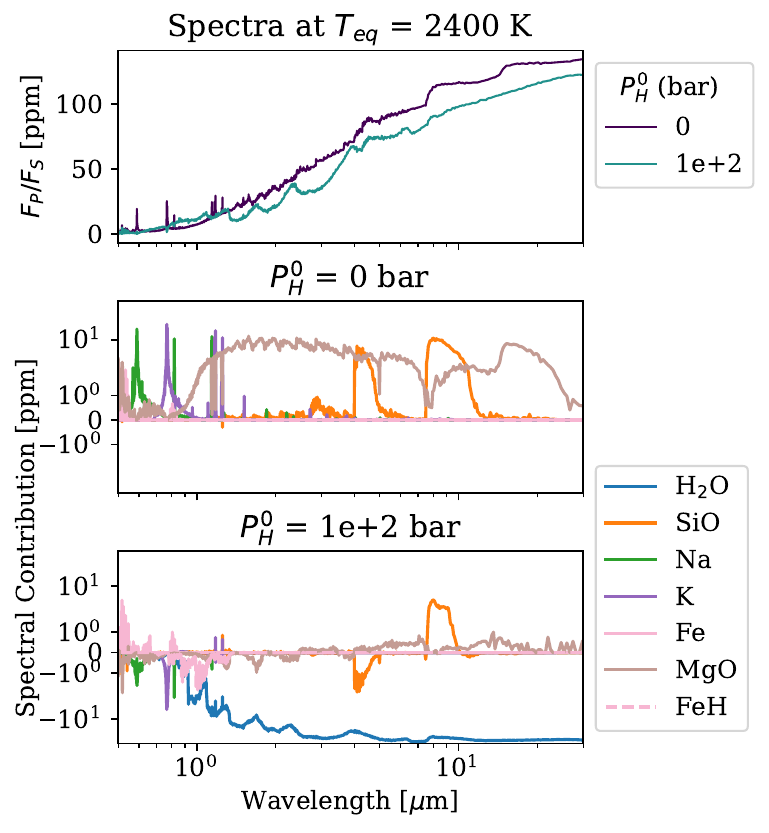}
  \caption{Emission spectra (planetary to stellar flux) and spectral contributions of molecules for an equilibrium temperature \Teq of 2400~K and for a hydrogen-free and a hydrogen-rich case (i.e., \ph~=~0 and $10^2$~bar).
  }
  \label{Fig_contribs_2400}%
\end{figure}

We have calculated the synthetic emission spectra corresponding to the above temperature and chemistry profiles.
The opacities that have been used are detailed in \app{sec:opacity_sources}.
The spectral resolution is equal to 5000.
The resulting spectra are shown in \fig{Fig_spectra_phs}.
The detailed spectral features for each species are also shown in \app{sec:contributions}.
The pressure-temperature profile and the chemistry have strong effects on the spectral features.
Thermal inversion is linked to emission features for the species that are present in the corresponding pressure range.
This is most notably visible for SiO, around 9~$\mu m$ (a feature already identified by \cite{ito2015,zilinskas2022}), but also for other species such as Na, K, Fe, and in some cases, MgO.
Indeed, the feature of MgO is most prominent around 11~$\mu m$, in the silicate case at \Teq$ = 2800$~K. 
Our model finds more MgO features than \cite{ito2015,zilinskas2022}) due to the outgassing of the magma ocean, as discussed in \app{sec:silicate_case}.
MgO features are also present between 1 and 3 $\mu m$.
These emission features tend to disappear for lower \Teq (i.e., longer orbital period) and when adding hydrogen.
The spectral contributions of each molecule for a hydrogen-free case and a hydrogen-rich case are shown in \fig{Fig_contribs_2400}.
We can see how most spectral emitting features are reduced or even absorbing in the hydrogen-rich case.
Interestingly, Fe is one of the only species that retains its spectral features when the hydrogen content is increased. This is more readily visible in 
\fig{Fig_contribs_2400} and \figures{Fig_contribs_0}{Fig_contribs_1e+2}.
Note that some species (in addition to H) are expected to escape to space over time, most readily Na and K \citep{charnoz2023effect}, a small study has therefore been led in \app{escape_na_k} to investigate atmospheres without these two species.

Some features, such as SiO at 4 $\mu m$, are linked to an emission peak or an absorption pit, following the PT profile. The feature is an emission peak for thermal inversions (i.e., for either low \ph or high \Teq), and an absorption pit in the presence of hydrogen due to the non-inverted PT profile.
PT profiles that are non-inverted (i.e., that do not exhibit a thermal inversion) are linked to absorption features mainly from \HTO, along with absorption features from 
other species, such as the SiO feature aforementioned.
Even though \cite{zilinskas2023observability} have not considered the chemical equilibrium between the volatiles and the vapor from the magma ocean (see also \sect{sec:sum_equilibrium}), they obtain similar trends. 

\subsection{A case with a very high hydrogen budget}
\label{sec:sih4}

\begin{figure}[ht]\centering
  \begin{minipage}{.3\textwidth}\centering
  \includegraphics[width=\textwidth]{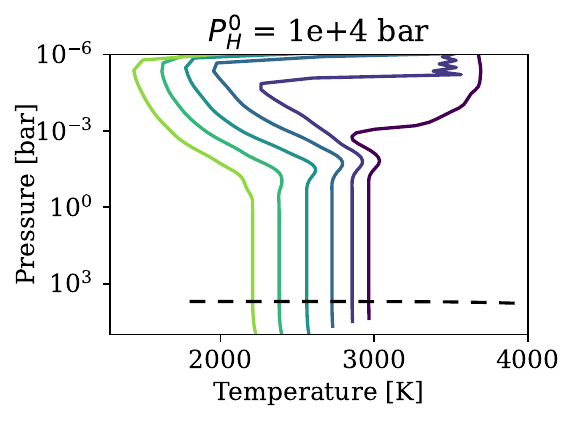}
  \end{minipage}
  \begin{minipage}{.17\textwidth}\centering
  \Teq 
  
  \includegraphics[width=\textwidth]{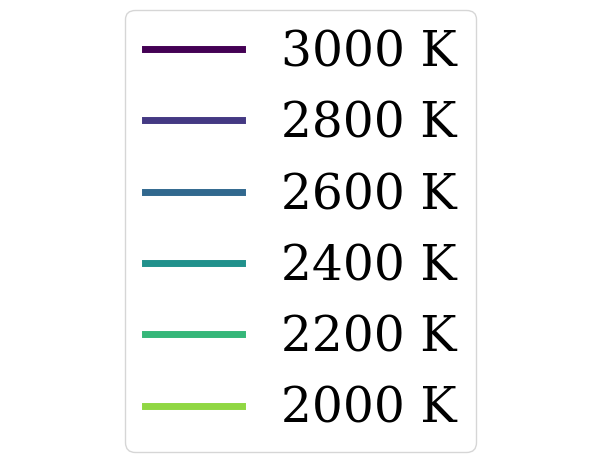}
  \end{minipage}
  \caption{Pressure-temperature profiles for a large hydrogen content \ph=$10^4$~bar, and equilibrium temperature \Teq ranging from 2000~K to 3000~K.
  }
  \label{Fig_pt_1e+4}%
\end{figure}

\begin{figure}[ht]\centering
\begin{minipage}{.29\textwidth}\centering
  \includegraphics[width=\textwidth]{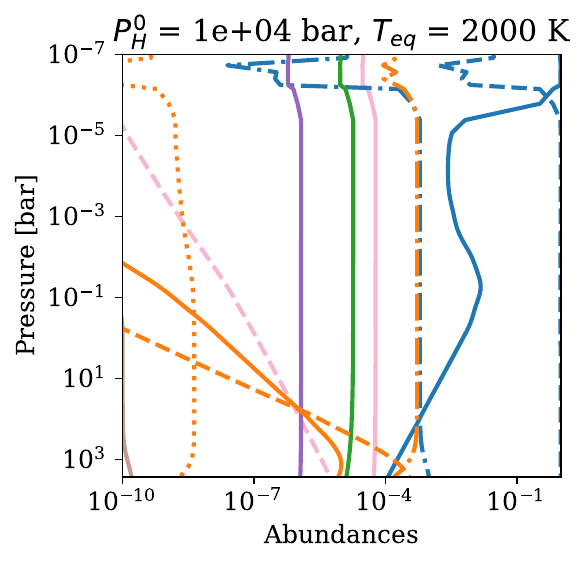}
  \includegraphics[width=\textwidth]{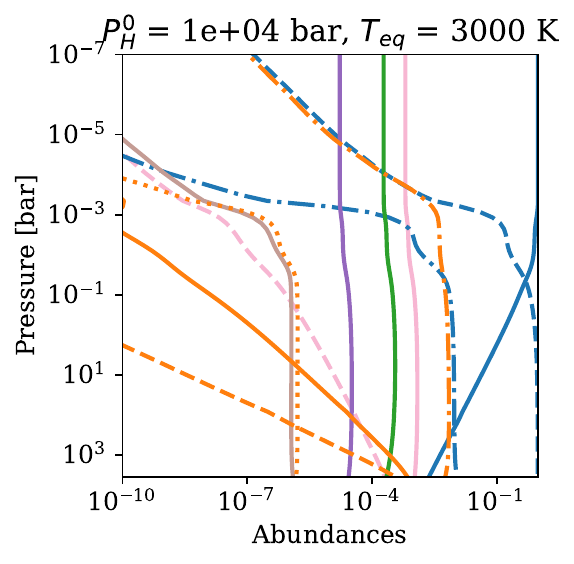}
\end{minipage}
\begin{minipage}{.1\textwidth}\centering
  \includegraphics[width=\textwidth]{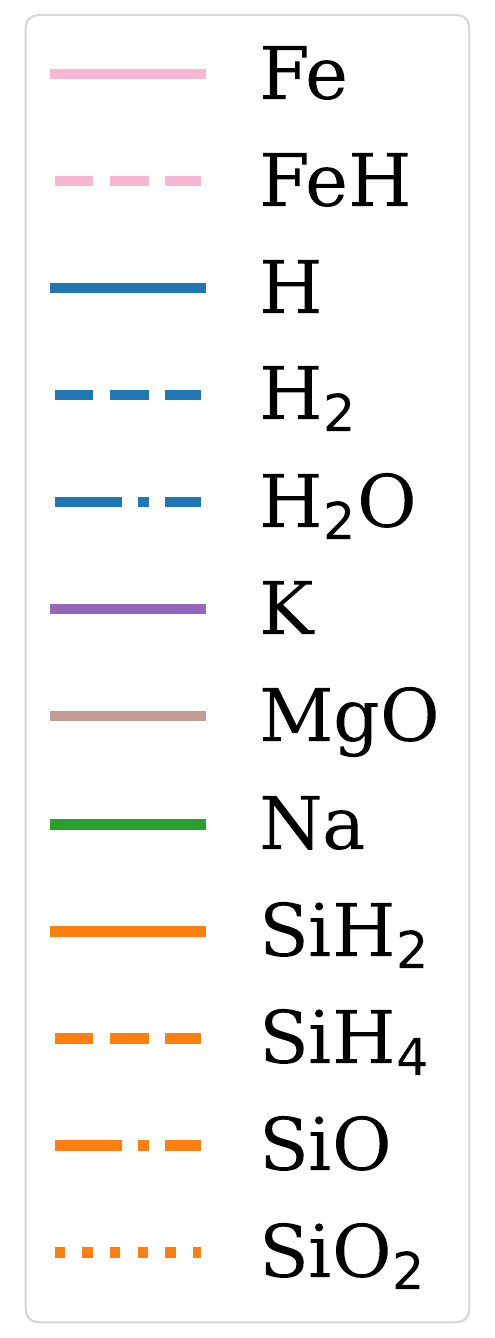}
\end{minipage}
  \caption{Abundances for very large hydrogen content $\ph~=~10^4$~bar, and \Teq~=~2000~K and \Teq~=~3000~K. Species also include \SiHF and \SiHt here, present at the surface but which dissociate at higher altitudes.
  }
  \label{Fig_x_1e+4}%
\end{figure}

\begin{figure}[ht]\centering
  \includegraphics[width=.5\textwidth]{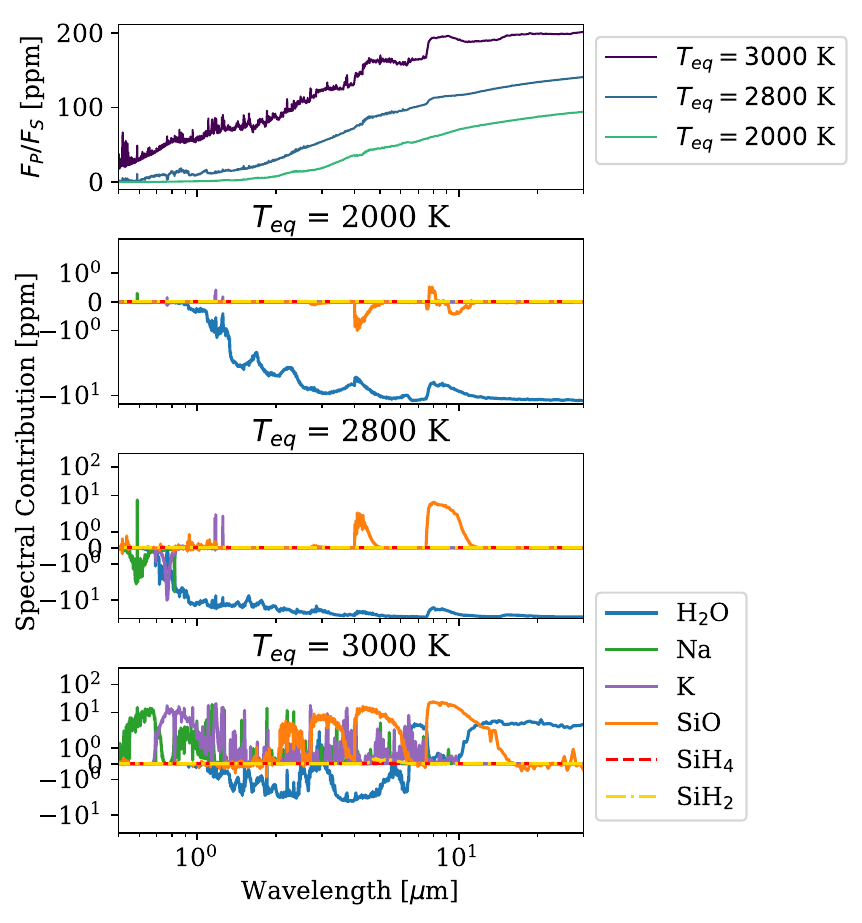}
  \caption{Spectral contributions of different species for \ph = $10^{4}$~bar of monoatomic hydrogen and for different equilibrium temperatures \Teq. \SiHF and \SiHt are negligible. \SiO is linked to very strong emission features at 2, 4, and 9 $\mu m$ at \Teq~=~3000~K.
  \HTO is also seen in emission at wavelengths$~>~10~\mu m$.
  }
  \label{Fig_contribs_1e+4}%
\end{figure}

We consider one final case with a very large amount of hydrogen in the atmosphere, and more precisely with \ph=$10^4$~bar of monoatomic hydrogen.
The pressure-temperature profiles are shown in \fig{Fig_pt_1e+4}.
This amount of hydrogen should account for no more than 1~\% of the total mass of an earth-like planet \citep{charnoz2023effect}, which is in line with the limits or predictions made by \cite{Elkins_2008,Dorn2021Hidden}.
The case at \Teq~=~3000~K shows a thermal inversion although there is a small non-inverted region around $10^{-3}$~bar. 
The remaining cases present a non-inverted profile that extends to the top of the atmosphere.
As shown by
Figs.~7~to~10 from \cite{charnoz2023effect}, 
the chemical composition of the vapor changes quite a lot from \ph~=~$10^2$~bar to $10^4$~bar.
Notably, new hydrogenated species appear
at \ph>$10^2$~bar, such as \SiHF, \SiHT and \SiHt.
They become dominant at \ph~>~$10^4$~bar.
However, as one can see in \fig{Fig_x_1e+4}, 
these species are only present in the lower part of the atmosphere, and disappear at high altitudes.
Consequently, they do not have any spectral contribution, as shown by \fig{Fig_contribs_1e+4} (see \fig{Fig_contribs_1e+4_6} for more details).
The atmosphere is completely dominated by \HT, \H being the second most dominant species (around 1\% at \Teq~=~2000~K).
Then \HTO and SiO are both present in almost equal quantities ($10^{-3}$ at 2000~K and $10^{-2}$ at 3000~K). 
At \Teq~=~3000~K, they both dissociate around $10^{-3}-10^{-4}$~bar.
SiO dissociates at higher altitudes and is therefore linked to stronger emission features due to the thermal inversion for pressures$~<~10^{-3}$~bar.
\HTO is mostly present in isothermal layers and its spectral (absorption) signature is therefore diminished, although it is linked to emission features for wavelengths$~>~10~\mu m$.

In this extreme case, only very hot planets will sustain a thermal inversion.
If we compare this case (\ph~=~$10^4$~bar of hydrogen) to the hydrogen-free case for \Teq~=~3000~K, we see that in both cases the temperatures reach 3700~K at high altitude.
However the surface temperatures of these two cases are quite different, as it is 2500~K in the hydrogen-free case and 3000~K for \ph~=~$10^4$~bar of hydrogen.

\subsection{Summary}

To summarize this section, 
\csname @beginparpenalty\endcsname10000
\begin{itemize}
    \item the presence of hydrogen tends to reduce the thermal inversion of short period super-Earths due to the formation of H$_2$O, watering down the silicate spectral features of such planets;
    \item the transition between a thermal inversion and a non-inverted profile is very sharp and sensible to the quantity of hydrogen;
    \item the surface temperature and pressure will be increased in the presence of hydrogen due to the greenhouse effect of \HTO;
    \item planets that are strongly irradiated will retain some emission features, due to the dissociation of water at higher and higher pressures;
    \item \SiHF and \SiHt (which form in very hydrogenated cases) are only present at the surface and have therefore no spectral contribution.
\end{itemize}

\section{Volatiles \& vapor: sum vs equilibrium}
\label{sec:sum_equilibrium}
\begin{figure*}
    \centering
   \includegraphics[width=\textwidth]{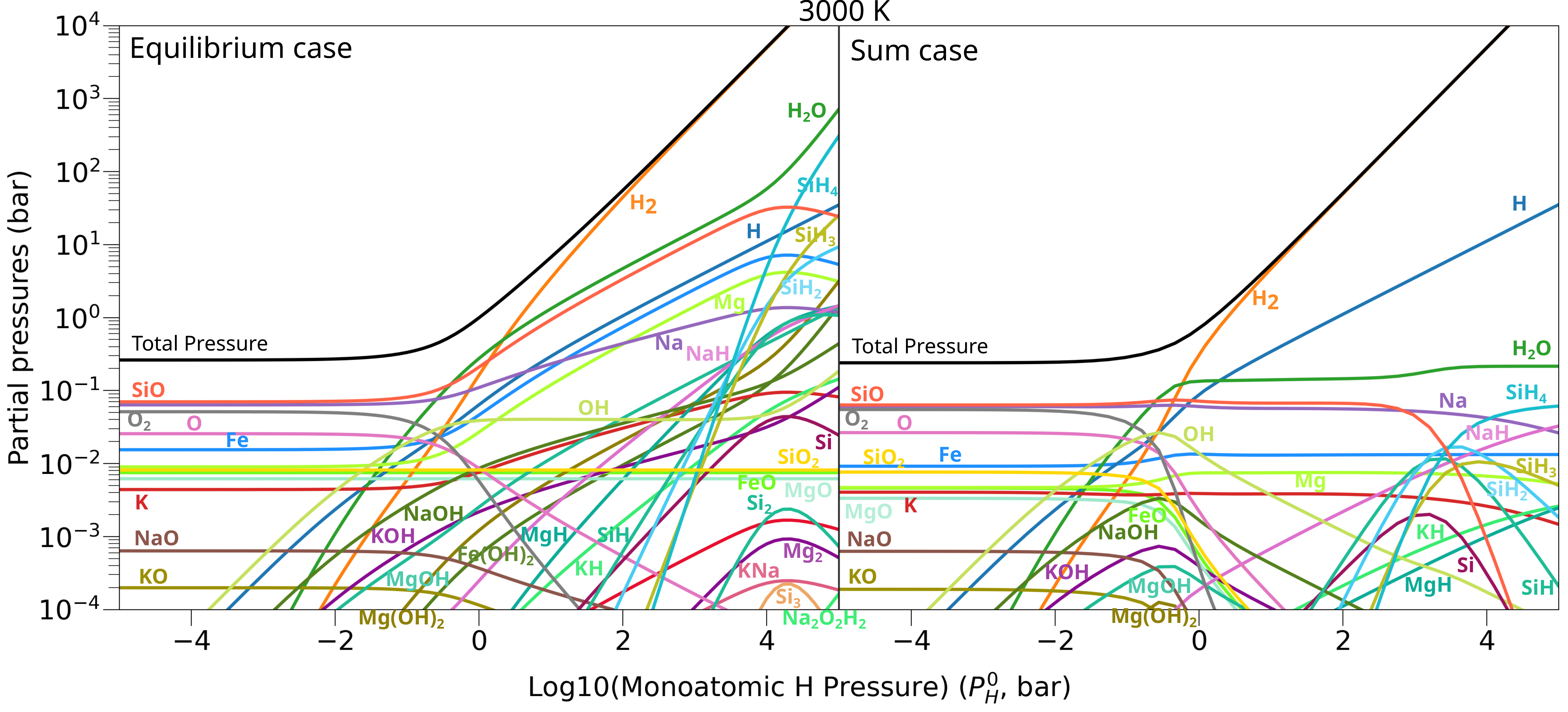}
    \caption{Partial pressures of most abundant species in the vapor calculated using the \Equilibrium method (left plot, same as Fig~B.3 from \cite{charnoz2023effect})
    and the \Sum method (right plot, methodology of \cite{zilinskas2023observability,piette2023rocky}), for different monoatomic hydrogen pressure \ph when the atmosphere is at equilibrium with the magma ocean at T~=~3000~K and hydrogen is \textbf{not} in equilibrium with the magma ocean.
    }
    \label{fig:compo_sum}
\end{figure*}

To compute the atmospheric composition, we have proposed in \cite{charnoz2023effect} and in the present study the computation of a gas-liquid equilibrium, in which hydrogen is in equilibrium with the gases evaporated from the magma ocean. 
As described in \cite{charnoz2023effect}, the set of gas-liquid equilibrium reactions and gas-gas reactions are solved simultaneously (including species released by the magma ocean + atmospheric volatiles), ensuring a full equilibrium but at the cost of a relatively long computation time. 
However, some studies \citep{zilinskas2023observability,piette2023rocky} follow a different approach (as an approximation of the full equilibrium): they sum the atomic abundances resulting from the evaporation of the magma ocean (in the absence of atmospheric volatile species) with the atmospheric volatile species (H,O, etc..). Then the gas-gas equilibrium of the mixture is computed with a gas-equilibrium code (like FastChem). The latter approach allows a faster computation, but cannot be considered as an exact gas-liquid thermodynamical equilibrium. 
We study here the differences between the two methods and evaluate the impact of the latter method onto the chemistry of the atmosphere, the associated pressure-temperature profile and emission spectrum.
The method in which the abundances of the vapor is summed to the volatiles \citep{zilinskas2023observability,piette2023rocky} will be referred to as the \Sum method, while the gas-liquid equilibrium discussed in this study and proposed by \cite{charnoz2023effect} will be referred to as the \textbf{Equilibrium} method.
The \Sum method includes hydrogen compounds but does not include other volatiles such as Carbon (while it is the case in \cite{zilinskas2023observability,piette2023rocky}), in order to be able to compare it to the \Equilibrium method proposed here, which only includes hydrogen as a volatile in the present study.
Both methods are computed via the \magmav code.

\fig{fig:compo_sum} shows the composition of the atmosphere at the surface using the \Equilibrium method (same as Fig.~B.3 from \cite{charnoz2023effect}),
and the \Sum method, when the atomic abundances of volatiles (H) are summed to the atomic abundances of the vapor.
Without computing the gas-liquid equilibrium in a self-consistent approach, the chemical composition of the atmosphere is quite different.
Overall, for a high hydrogen content (\ph>1~bar), the \Sum method will underestimate the partial pressures of the gases outgassed by the magma ocean in the final mix, compared to the \Equilibrium method.
This holds as well for \HTO.

\begin{figure}[ht]\centering
  \includegraphics[width=.5\textwidth]{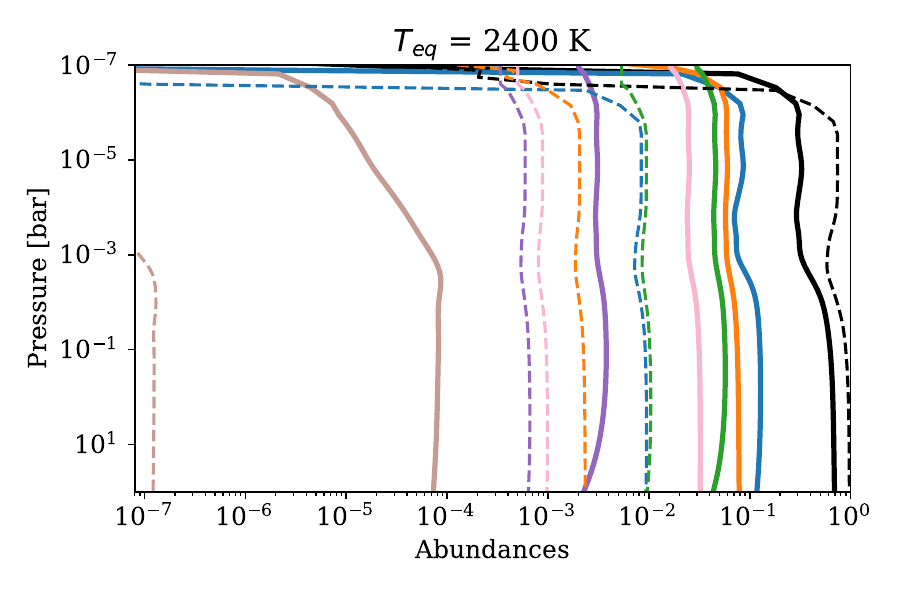}
  \begin{minipage}{.3\textwidth}
\includegraphics[width=\textwidth,height=.1\textheight,keepaspectratio]{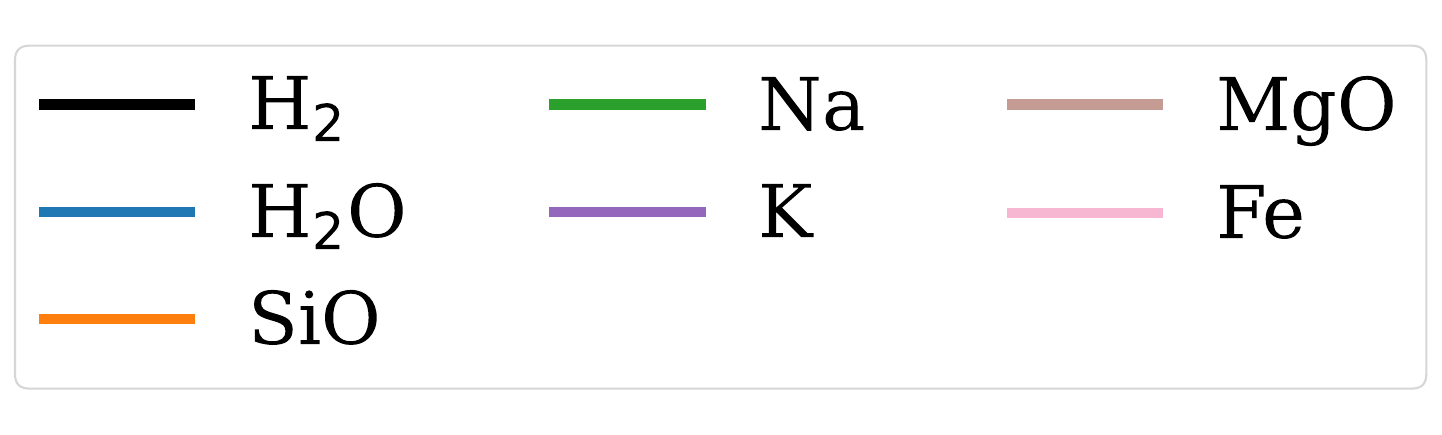}
\end{minipage}
\begin{minipage}{.17\textwidth}
\includegraphics[width=\textwidth,height=.1\textheight,keepaspectratio]{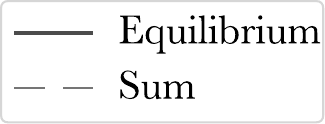}
\end{minipage}
  \caption{Abundances for a case with some hydrogen (\ph~=~$1$~bar), at \Teq~=~2400~K.
   \Sum is the case where volatiles are summed to the vapor content.
  \Equilibrium is the case where the gas-liquid equilibrium is computed.
  }
  \label{Fig_x_sum}%
\end{figure}

The atmospheric chemistry (shown in \fig{Fig_x_sum}) is therefore also quite different between the two methods.
The main aspects of these differences are twofold:
\begin{enumerate}
\interlinepenalty10000
    \item the abundances of \HTO and silicates (SiO, Na, K, Fe, MgO) are lower in the \Sum method compared to the \Equilibrium method by at least one order of magnitude;
    \item \HT is more abundant in the \Sum method.
\end{enumerate}
Some species (such as FeH, H, ...) have not been included in the graph as they have no spectral significance, although they are present in the model (see \fig{Fig_x} for example).

\begin{figure}[ht]\centering
  \includegraphics[width=.4\textwidth]{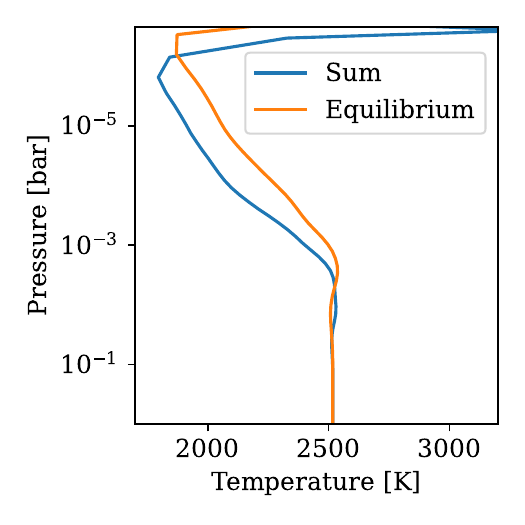}
  \caption{Thermal structure of the atmosphere calculated by \atmo for \Teq~=~2400~K and \ph~=1~bar.
  \Sum is the case where volatiles are summed to the vapor content.
  \Equilibrium is the case where the gas-liquid equilibrium is computed.
  }
  \label{Fig_pts_sum}%
\end{figure}
This leads to a difference of a few hundred Kelvin degrees for pressures below $10^{-3}$~bar, the \Equilibrium case being the hottest, as shown by \fig{Fig_pts_sum}.
The surface temperature does not seem to be impacted.
The thermal inversion is pushed to higher altitudes in the \Equilibrium case.

\begin{figure}[ht]\centering
  \includegraphics[width=.4\textwidth]{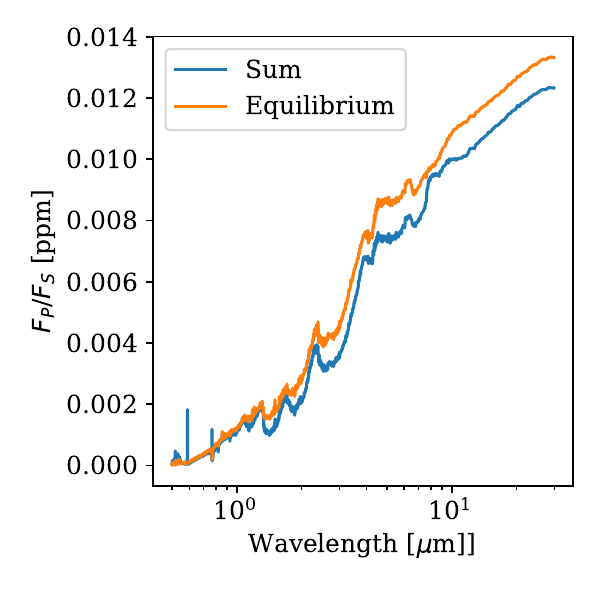}
  \caption{Emission spectra (planetary to stellar flux) for \Teq~=~2400~K and \ph~=~1~bar.
  \Sum is the case where volatiles are summed to the vapor content.
  \Equilibrium is the case where the gas-liquid equilibrium is computed.
  }
  \label{Fig_spec_sum}%
\end{figure}

Due to its lower temperature, the emission spectrum of the \Sum case is also lower than that of the \Equilibrium case.
This is shown in \fig{Fig_spec_sum}.
The spectra also exhibit different spectral features.
We can see Na as an emission feature for example in the \Sum method, while it is invisible in the \Equilibrium method. 
The 9~$\mu m$ SiO feature is visible in emission in the \Sum method while it is absent (and slightly absorbing) in the \Equilibrium method.
The same can be said for the Na and K features between 0.6 and 0.8 $\mu m$.
One can refer to \sect{sec_contrib_sum} for more information on the exact spectral features contributing to these two spectra. 

To conclude this section, we would like to stress that taking into account volatiles in the gas-liquid equilibrium changes fundamentally the whole atmospheric structure.

\section{Discussion}


\subsection{Observational target candidates}

In this study, we would like to focus on targets that match the criteria necessary for maintaining a magma ocean on the day-side of the planet and of which the radius is sufficiently large
for us to consider a potential (thin) hydrogen envelope,
without falling into the category of gas giants.
To that end, we have developed a hydrogenation potential index that will be discussed in \sect{sec_hydrogen_index}.

\begin{figure*}[ht]
    \centering
    \includegraphics[width=.9\textwidth]{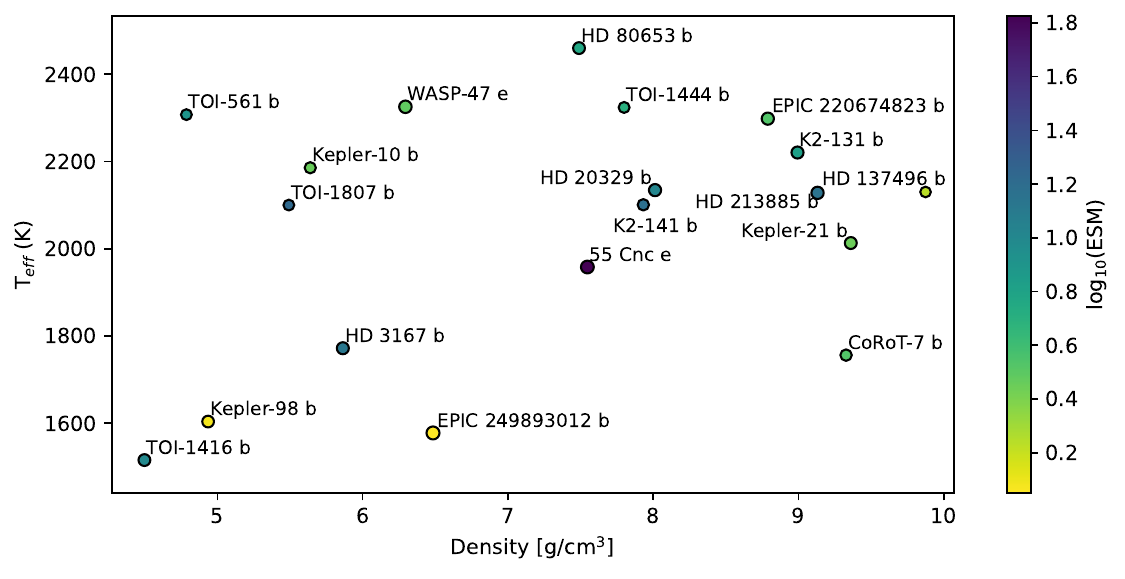}
    \includegraphics[width=.9\textwidth]{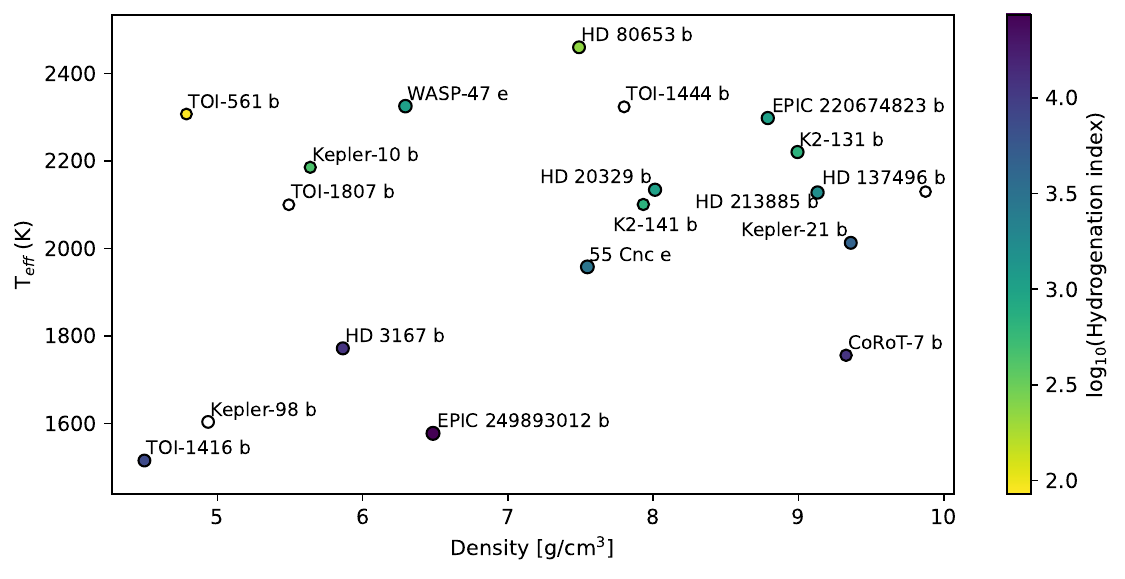}
    \caption{Subset of super-Earths of interest (with a better chance at being hydrogenated), displayed by their density and equilibrium temperature \Teq. 
    The size indicates the planet radius.
    Top plot: the color indicates the planet Emission Spectroscopy Metric (ESM) \citep{kempton2018esm};
    the planets with the highest \Teq are the best candidates for magma oceans planets.
    Bottom plot: the color indicates the hydrogenation index $\chi$ (see \sect{sec_hydrogen_index});
    all colored planets are good candidates for hydrogenated planets ($\chi~>~1$), the others are not colored.
    Source: \url{http://exoplanetarchive.ipac.caltech.edu}}
    \label{fig:rad_orb}
\end{figure*}

\fig{fig:rad_orb} shows a list of potential targets of interest.
The planets with the highest Emission Spectroscopy Metric (ESM) \citep{kempton2018esm} are the best candidates for observation.

Only a limited number of targets match the criteria for being characterized as hot rocky super Earths, with an equilibrium temperature sufficiently high to be capable of sustaining a magma ocean, 
and simultaneously have the necessary size and low density to also consider a potential hydrogen envelope.

\subsection{The Maximum Hydrogenation index: $\chi$}
\label{sec_hydrogen_index}

In addition to the equilibrium temperature we need another selection criterion.
Our objective is to provide a qualitative assessment of whether a given planet (among the high temperature rocky exoplanets displayed in \fig{fig:rad_orb}) could potentially harbor a hydrogen layer above a magma ocean considering its measured mass and radius.
This is not a simple question as the more hydrogen, the lower the average density of the planet,
and the thicker the atmosphere.
Indeed, as we add hydrogen, we increase the temperature at the bottom of the atmosphere due to the greenhouse effect of \HTO, and decrease the mean molar mass $m$, thus increasing the scale-height of the atmosphere, as the scale-height  $H~=~RT/mg$,
with R standing for the ideal gas constant, $g$ the surface gravity and T for the atmosphere temperature.
We try to estimate the maximum hydrogen content of the planet.
For that, we consider a simple 2-layers planet model and test many atmospheric compositions to determine the maximum amount of hydrogen that can be put in the atmosphere while matching its measured planet mass and radius.

Let us consider a planet with measured mass \Mp and measured radius \Rp, with average density $\rho_p$, and equilibrium temperature $T_p$
(here we simply use the published $T_p$ value in the discovery paper).
The planet is considered to consist of 2  layers: a ``solid'' planet with mass \Mc and radius \Rc surrounded by an atmosphere with thickness Z (Z~$<<~$\Rc). Z here is the 
distance between the solid surface and where the atmosphere becomes optically thin (low opacity),
so that the equivalent transit radius of the planet is $\Rp=\Rc+Z$. If the planet has no atmosphere (which we do not know) then we have \Rp~=~\Rc, \Mp~=~\Mc and Z~=~0. 
\Rc and \Mc are assumed to be constants (i.e. independent of atmospheric composition), as we do not account for the dissolution of H in the magma ocean.
\cite{Dorn2021Hidden} indeed report that the density variation of the mantle is by about $-1\%$ for a $1\%$ mass fraction of \HTO dissolved in the magma ocean \citep{Bajgain2015}. This variation is very small compared to all other approximated quantities in this simple model so we have not taken that into account.

If we assume some atmospheric composition with mean molar mass
$m$ and surface pressure $P_0$ ($m$ and $P_0$ are taken from our \magmav model, see \sect{sec:vaporization_model})  we want the total mass of the planet to be equal to \Mp. 
For a thin atmosphere, the atmosphere mass $M_a$ is related to $P_0$ through:
\begin{equation}
P_0=\frac{M_a g}{(4\pi \Rc^2)},
\end{equation}
\begin{equation}
g=\frac{GM_c}{\Rc^2},
\end{equation}
$G$ is the gravity constant.
The solid planet average density $\rho_c$ is:
\begin{equation}
\rho_c~=~\frac{M_c}{\frac{4}{3}\pi\Rc^3},
\end{equation}
so we get:
\begin{equation}    
M_a=\frac{3 P_0 \Rc}{\rho_c G}.
\end{equation}
Since $M_c+M_a=M_p$, we have:

\begin{equation}
\frac{4}{3}\pi \Rc^3\rho_c+\frac{3 P_0 \Rc}{\rho_c G}=\Mp.
\label{eq: Equ_H_Rp}
\end{equation}

For a simple hydrostatic and isothermal atmosphere we have $P(z)=P_0 e^{-zmg/RT}$.
We set the pressure at which the atmosphere becomes transparent to be about 0.01 bar (=~$P_t$), 
following \fig{Fig_optical_depth}. 
This approximation should be taken with caution and needs to be refined in the future.
We have
Z~=~$-(RT/mg) \ln(P_t/P_0)$. Since $Z+\Rc=\Rp$, the following equation must be verified:

\begin{equation}
-(RT/mg) \ln(P_t/P_0)+\Rc=\Rp.
\label{eq: Equ_H_Mp}
\end{equation}

In equations \ref{eq: Equ_H_Rp} and \ref{eq: Equ_H_Mp}, \Mp and \Rp and T are observational data, m and $P_0$ are the 
mean molar mass and pressure of the atmosphere (considered to be free parameters), and \Rc and \Mc are the unknowns (that will depend on the choice of $m$ and $P_0$). 
$P_0$ and $m$ are chosen among the compositions given by our \magmav code at temperature $T_p$ and for various amount of hydrogen $P_H^0$. For a given temperature $T_p$ we determine \Rc and \Mc for all atmospheric composition ($m$) with $P_H^0$ ranging from 0 to $10^5$ bar. To stay in the regime of rocky-exoplanet or mini-neptune, we will consider as ``valid'' those solutions for which Z$~<<~\Rc$ and those for which the $\rho_c<1.2~\rho_p$ (so that the solid planet average density is at maximum 20$\%$ higher than the planet's average density). 
These choices are arbitrary but they ensure that our 2 layers planet stays in the range of ``rocky'' world or ``mini-neptune''. So for each planet we determine the maximum quantity of hydrogen, $P_H^0$ we can put in the atmosphere, while matching the observations and the constrains above. 
It has been shown in \cite{charnoz2023effect} that for a magma-ocean planet to be considered as ``hydrogenated'' (so that it contains lots of \HTO, \HT in addition to Na, SiO, MgO, etc.) the hydrogen pressure ($P_H^0$) must be larger than \Pmineral (the pure-mineral atmosphere pressure) fitted by Equation \eq{eq_pmineral}.
Considering this finding, we define our Maximum Hydrogenation index as $\chi~=~P_H^0/\Pmineral(T)$, in other words, this is the maximum hydrogen pressure we can put in the planet's atmosphere, divided by the pressure of a pure mineral atmosphere at the same temperature.
If $\chi~<~1$ then the planet can only have a pure mineral atmosphere, dominated by Na, SiO, MgO, etc. If $\chi~>~1$ the atmosphere can potentially bear significant H and molecules like \HT, \HTO (in addition to Na, SiO, MgO, etc.). Of course, the larger $\chi$ the larger the \HT and \HTO abundances (and all other hydrogenated species).

The Maximum Hydrogenation Index ($\chi$) is color plotted in \fig{fig:rad_orb} (bottom plot) for planets with $T_p~>~1500~$K. We see that lower temperature planets have a higher $\chi$: this is because low temperature planets have a low \Pmineral, and thus are more easily hydrogenated. Planets with larger $g$ (in general larger planets) have also a high  $\chi$ because their $g$ can accept more massive atmosphere with small change in their transit radius. Interestingly all planets present in \fig{fig:rad_orb} (bottom plot) could have a hydrogenated atmosphere ($\chi~>~1$) while being molten. This figure may be useful to select planets that may be potentially molten while accepting a significant fraction of H.

\subsection{Caveats}
\label{sec:discussion}

Some effects have not been taken into account which could be of some importance for future studies.

Indeed, this study does not include the dissolution or the exsolution of \HTO in/from the magma ocean.
We have assumed that H is already present in the atmosphere and does not exsolve from the magma ocean.
The exsolution of \HTO will lead to a more oxidized atmosphere according to our preliminary investigations (not included in the current study), in which the atmosphere is dominated by \HTO rather than \HT.
Moreover, according to \cite{Kite_2020}, atmospheric \HTO/\HT ratio is proportional to magma FeO content.
Concurrently, \cite{bower2022,Maurice2023arXiv230107505M} have also shown that the \HTO/\HT ratio increases when the magma ocean solidifies.
The change is quite drastic, as the atmosphere can be dominated by \HT in the early stages of the planet formation and rather dominated by water in later stages. This also depends on the proximity of the planet with the star, since the magma ocean is also expected to survive longer under intense stellar irradiation.
Taking into account the dissolution of 
\HTO will change the activity coefficients of the melt and therefore also change the atmospheric composition.
Overall, the main effect of the dissolution of \HTO to be expected is an increase in the abundance of \HTO in the atmosphere, and thus in the spectral signature of \HTO.

The inclusion of carbon, as well as other volatiles, in the system should also be considered for future studies, as this should have a considerable impact on the atmospheric structure and chemistry. 
Indeed, as shown by \sect{sec:sum_equilibrium}, taking into account hydrogen in the gas-liquid equilibrium changes drastically the chemistry of the atmosphere, in part due to its interactions with oxygen.
Other species that also interact with oxygen may have a similar effect on the overall chemistry.

This study does not focus on atmospheric escape (although we have partially discussed this in \cite{charnoz2023effect}).
H is expected to escape quite fast in smaller bodies.
Its effect on the chemistry of the evaporated species is also promoting the escape to space of heavy elements such as Na and K.
As discussed in \app{escape_na_k}, the escape of Na and K is not fundamentally changing the characteristics of the atmospheric structure.
The impact of the escape of H is the major aspect that should be taken into account in future studies.
Young planets are expected to have retained more H than older planets.
This study does not focus on either, although the content of H that we have used as input (\ph) could be constrained using parameters such as the age of the planet and its proximity to its host star.

The present study follows a 1D approach to study the day-side thermal emission.
However, as shown by
\cite{zieba2022} for K2-141~b for example,
two-dimensional models better explain the data of hot rocky exoplanets.
\cite{Kite_2016} show that for planets with a substellar temperature above 2400~K, the magma overturning circulation is slow compared to atmospheric transport.
\cite{castan2011atmospheres} also point out that rotation could break the substellar-point symmetry in a 2D equatorial simulation and,
under conditions of permanent hemispheric forcing, could lead to the
formation of superrotating equatorial winds \citep{Showman_2011}.
This is all a strong incentive to use a 2D atmospheric structure in order to model the winds and heat transport to the night-side.

In the context of gas giants, \cite{tremblin2017advection} also discuss the effect of the two-dimensionality of the atmospheric model on the pressure-temperature profile of strongly irradiated planets.
They show how the 1D model follows an isothermal profile around 1~bar, while the 2D model becomes adiabatic, which could explain their inflated radius.
This adiabatic profile is forced by the deep circulation forced by the asymmetric irradiation that transports energy downwards in the atmosphere. This becomes the main energy transport as soon as the atmosphere is sufficiently optically thick so that radiation is inefficient to transport energy. In a relatively thick \HT/He atmosphere of a rocky planet reaching e.g. hundreds of bars at the surface, this process should also take place and also lead to a much higher surface temperature compared to a 1D model ignoring circulation processes.
This may also be applicable to the present study, and our 1D model is thus likely underestimating the actual scale-height of the atmosphere.

\section{Conclusion}

We have combined a vaporization model with an atmospheric model (\magmav and ATMO) to compute the thermal and chemical structure of the atmosphere of a molten rocky planet in the presence of a hydrogen layer. Our study focuses on the effect of H on the atmospheric structure and emission spectrum.

In a pure silicate atmosphere, an atmospheric thermal inversion occurs, due to absorption in the visible range, from Na, K, Fe, SiO, etc.
We find the same emission peaks of SiO at 9~$\mu m$ as \cite{ito2015,zilinskas2022}.

Our Na-K-Mg-Al-Fe-Si-O+H model (\magmav), also accounts for the presence of hydrogen in the atmosphere in the calculation of the gas-liquid equilibrium.
Our findings are that hydrogen drastically changes the atmospheric composition, and thus changes the atmospheric structure.
For atmospheres relatively ``cool'' (<~2500~K), a low hydrogen budget is enough to remove the thermal inversion as water becomes dominant, and molecules will dissociate only in the higher levels of the atmosphere, at low pressure.
For hotter atmospheres (>~2500~K), molecules start to dissociate at higher pressures ($10^{-3}$ to $10^{-5}$ bar), and a higher hydrogen budget is not enough to remove the thermal inversion (as water is dissociated).
This confirms previous trends observed by \cite{zilinskas2023observability} as to the removal of thermal inversion due to water in the atmosphere.

Two distinct regimes appear, the transition from one state to the other being very sharp.
At high hydrogen content, 
H forms water, and the resulting greenhouse effect increases the temperature at the surface
of the magma ocean well above the equilibrium temperature by several hundred Kelvin degrees 
(\fig{Fig_tsurf}).
The effect is less strong for higher equilibrium temperatures.
At low hydrogen content, metallic species absorbing in the visible create a thermal inversion in the upper atmosphere. In that case the surface temperature is in general lower than the equilibrium temperature (for T>2200 K). 
This effect is less strong for lower equilibrium temperatures.

What are the spectral features that we could expect from a hydrogenated magma ocean planet?
In the absence (or very low abundance) of hydrogen, thermal inversion induces emission features from SiO, notably at 9~$\mu m$, but also MgO, Na, K and Fe. When the hydrogen content increases, 
the stronger greenhouse effect caused by the increase in water vapour induces
a non-inverted pressure-temperature profile, and spectral absorption features, mainly from \HTO, but also of SiO. The emission peak of SiO around 4~$\mu m$ switches to absorption for a higher hydrogen budget.
The emission features of other species is also reduced.
Fe is the only species that retains a quite strong emission feature, with or without the addition of hydrogen. At high hydrogen content, \SiHF is produced at the base of the atmosphere, near the surface, but disappears from the upper layers and is therefore not linked to any spectral features. This holds as well for \SiHt.

Future studies should focus on the inclusion of other volatiles in the calculation of the gas-liquid equilibrium, as it might prove to have a major influence over the whole atmospheric structure and chemistry and have quite different spectral features. 

Finally we have introduced and discussed a criterion, a maximum hydrogenation index, which gives an insight as to the potential of a planet to host a hydrogenated atmosphere.

\begin{acknowledgements}
SC, AF, PT acknowledge financial support by LabEx UnivEarthS (ANR-10-LABX-0023 and ANR-18-IDEX-0001) and by the CNES (Centre National d'Études Spatiales). PT would also like to acknowledge and thank the ERC for funding this work under the Horizon 2020 program project ATMO (ID: 757858). 
\end{acknowledgements}

\bibliographystyle{aa}
\bibliography{biblio}

\appendix

\section{Pure silicate case}
\label{sec:silicate_case}

Silicate atmospheres (i.e., hydrogen-free in our context) have been shown to exhibit a strong thermal inversion \citep{ito2015,zilinskas2022}, linked to an emission feature of SiO around 9~$\mu m$.
We here confirm these results, and compare the vapor content calculated via \magmav to the LavAtmos code \citep{van2022lavatmos}, discussing the differences produced on the resulting atmospheric structure and emission spectrum.

\begin{figure}[ht]\centering
  \includegraphics[width=.5\textwidth]{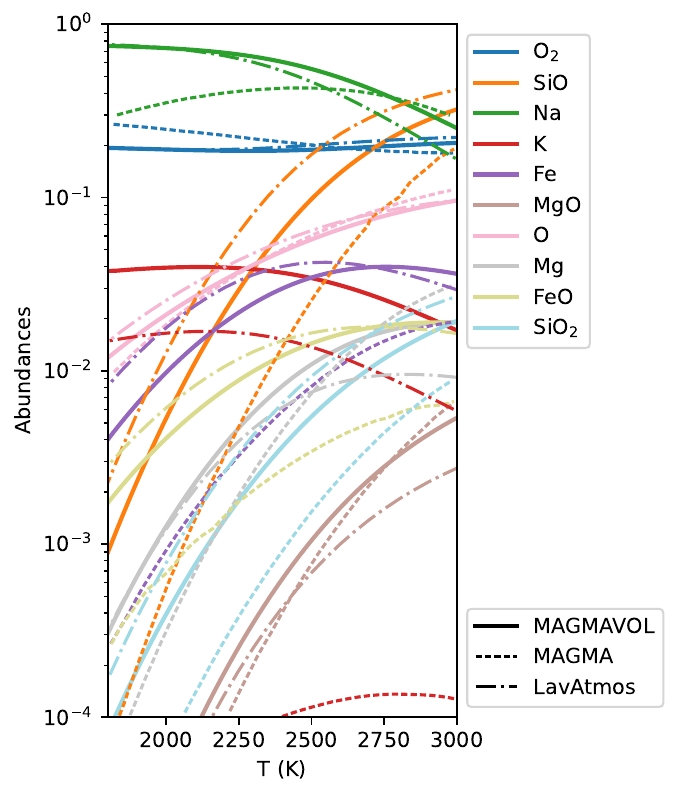}
  \caption{Comparison between the composition of the vapor calculated by our vaporization code, 
  in solid lines, and MAGMA, in dashed lines \citep{Schaefer_2009},
  and LavAtmos, in dash-dotted lines \citep{van2022lavatmos}. 
  There is no Hydrogen in this case. LavAtmos includes Al, Ca and Ti, that are not present in our code.}
  \label{fig:element_abundances_lavatmos}%
\end{figure}

\fig{fig:element_abundances_lavatmos} shows a comparison between the composition of the vapor calculated by our model (\magmav, \cite{charnoz2023effect}), MAGMA \citep{Schaefer_2009} and
LavAtmos \citep{van2022lavatmos,zilinskas2022,zilinskas2023observability} for the silicate-only case (\ph = 0).
Our model predicts similar abundances as LavAtmos for Na, \OT, O and Fe, but predicts less SiO (by a factor of $\sim$2) and more K (factor of 3).
MAGMA predicts less Na at low temperatures, and less SiO and Fe ($\sim$1 order of magnitude) and much less K (two orders of magnitude) than the two other models.
\cite{van2022lavatmos} explain the significant difference between the K abundances by the difference in calibration of the thermodynamic models.
Our model yields results that align more closely with LavAtmos than with MAGMA.

\begin{figure}[ht]\centering
  \includegraphics[width=.5\textwidth]{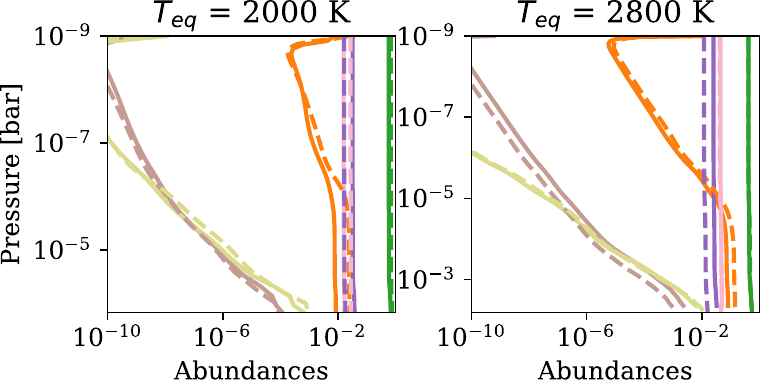}
\begin{minipage}{.32\textwidth}
\includegraphics[width=\textwidth,height=.1\textheight,keepaspectratio]{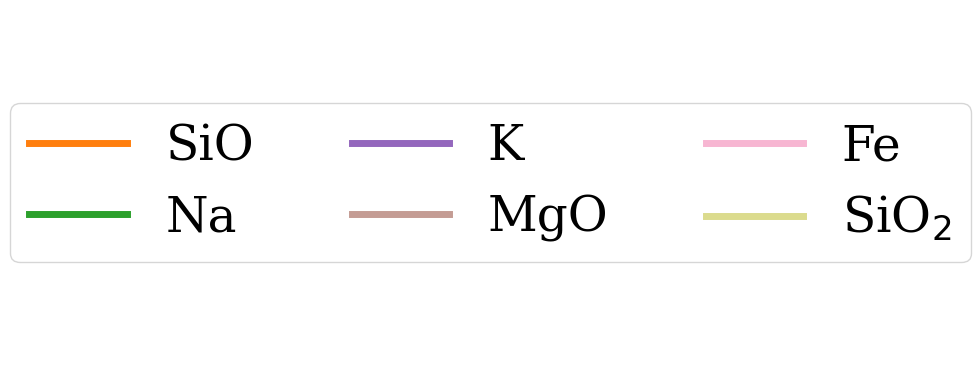}
\end{minipage}
\begin{minipage}{.15\textwidth}
\includegraphics[width=\textwidth,height=.09\textheight,keepaspectratio]{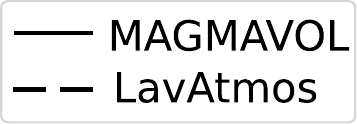}
\end{minipage}
  \caption{Abundances in a silicate atmosphere, computed via \atmo, using the vapor computed by \magmav and LavAtmos.
  There is no fundamental difference in the chemistry computed using the vapor from both codes.
  }
  \label{Fig_x_silicate}%
\end{figure}

The atmospheric chemistry for \magmav and LavAtmos 
is shown in \fig{Fig_x_silicate} for two equilibrium temperatures, \Teq~=~2000~K and \Teq~=~2800~K.
The two models still agree very well, although, as could be expected from \fig{fig:element_abundances_lavatmos}, SiO is slightly more abundant in the vapor computed by LavAtmos.

\begin{figure}[ht]\centering
  \includegraphics[width=.5\textwidth]{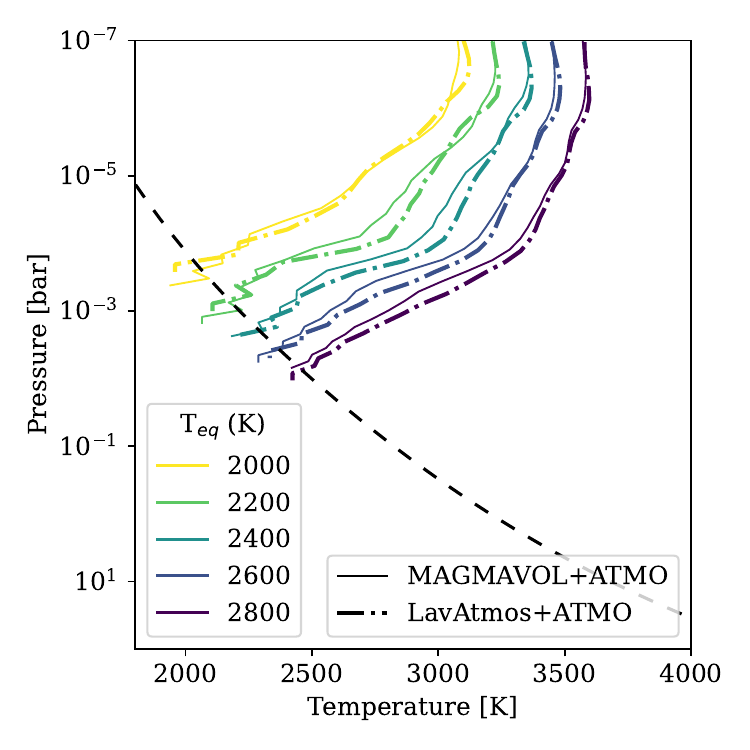}
  \caption{Thermal structure of the silicate atmosphere calculated by \atmo for different equilibrium temperatures \Teq (indicated by the colors), which translates as different orbital radii.
  The dash-dotted PT profiles use the vapor computed via LavAtmos, while the plain lines use \magmav.
  The dashed line show the vapor pressure \Pvap(T) corresponding to the limit between the magma ocean and the gaseous atmosphere.
  }
  \label{Fig_pts_silicate}%
\end{figure}

\fig{Fig_pts_silicate} shows the pressure-temperature profiles of atmospheres for which the equilibrium temperature \Teq is between 2000 and 3000~K in the case of a pure silicate atmosphere.
We cannot see major differences between the atmospheres calculated from the two vaporization models.

\begin{figure}[ht]\centering
  \includegraphics[width=.45\textwidth]{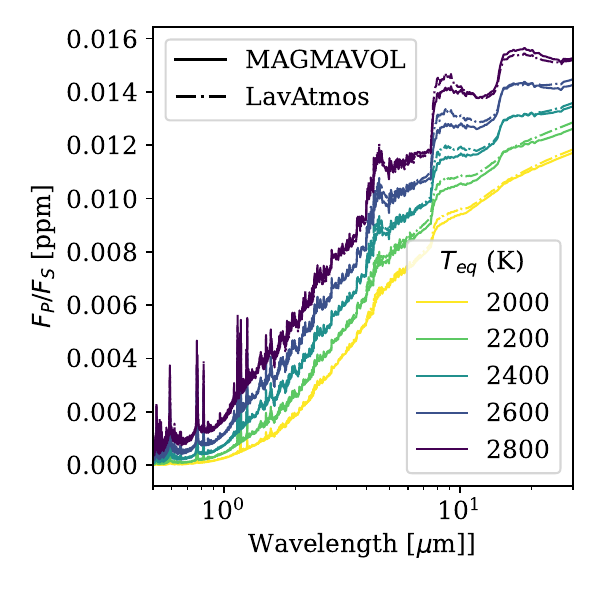}
  \caption{Emission spectra (planetary to stellar flux) for vaporizations calculated by LavAtmos and \magmav.
  The vapor from LavAtmos contains slightly more Si, which is noticeable in the SiO band around 9~$\mu m$. 
  }
  \label{Fig_spectra_silicate}%
\end{figure}

The slightly higher abundance of SiO of LavAtmos is visible in the simulated emission spectra, shown in \fig{Fig_spectra_silicate}.
On this figure, we can see that the baseline for the spectrum computed via \magmav is lower than that of LavAtmos, due to the fact that the corresponding temperature profiles are generally colder (\fig{Fig_pts_silicate}).
But we can see the increased emission in the SiO band around 9~$\mu m$ due to SiO being slightly more abundant in the LavAtmos model.
On the contrary, the features of MgO are reduced using LavAtmos.

\FloatBarrier
\section{Escape of Na and K}
\label{escape_na_k}

\begin{figure}[ht]\centering
  \includegraphics[width=.5\textwidth]{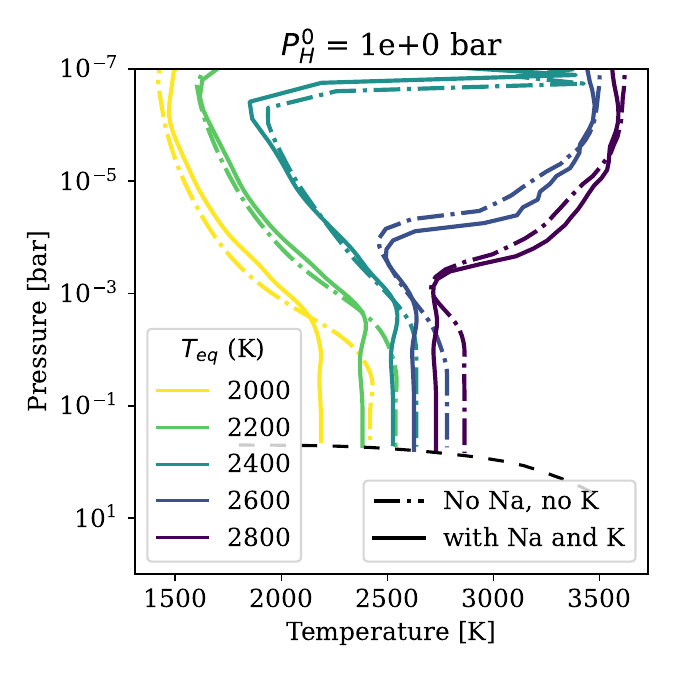}
  \caption{Thermal structure of the atmosphere calculated by \atmo for different equilibrium temperatures \Teq (indicated by the colors), i.e., different orbital radii.
  The partial pressure of monoatomic hydrogen \ph is set to 1~bar.
  The dash-dotted lines indicate the case without Na nor K.
  The dashed line show the vapor pressure \Pvap(T) corresponding to the limit between the magma ocean and the gaseous atmosphere.
  }
  \label{Fig_pts_no_na_no_k}
\end{figure}

As shown by \cite{charnoz2023effect}, Na and K would be among the first species (apart from H) to escape from the atmosphere.
\fig{Fig_pts_no_na_no_k} shows the difference between the atmospheric structure for a case with Na and K compared to a case where Na and K have disappeared, for which we recompute the gas-liquid equilibrium.
The upper atmosphere is not affected by this change.
However, the surface temperature is increased by more than 100~K in the absence of Na and K (and up to 200~K for lower \Teq).
This is not sufficient to have any spectral impact (except for Na and K features).

\section{Opacity sources}
\label{sec:opacity_sources}

\begin{figure*}[ht]\centering
  \includegraphics[width=\textwidth]{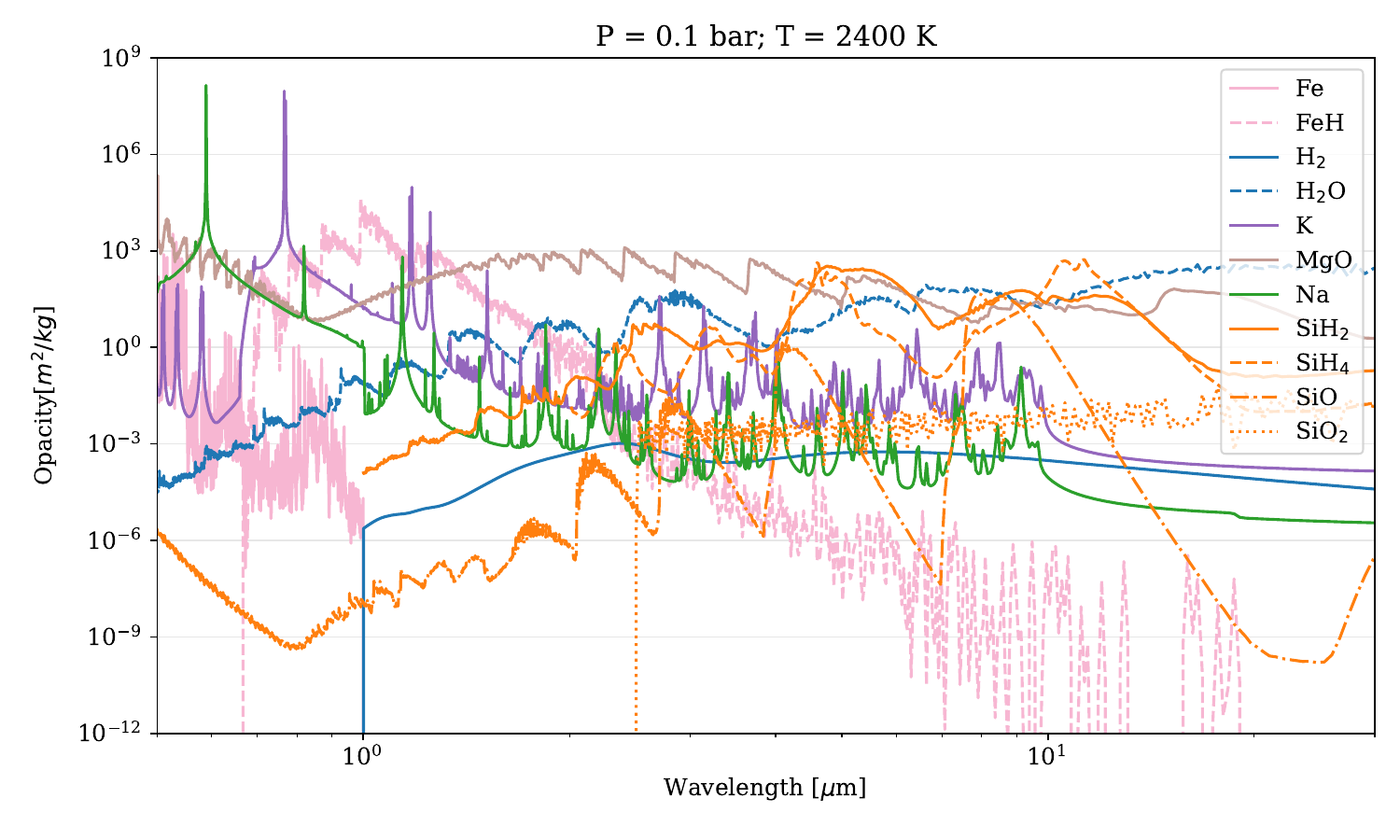}
  \includegraphics[width=\textwidth]{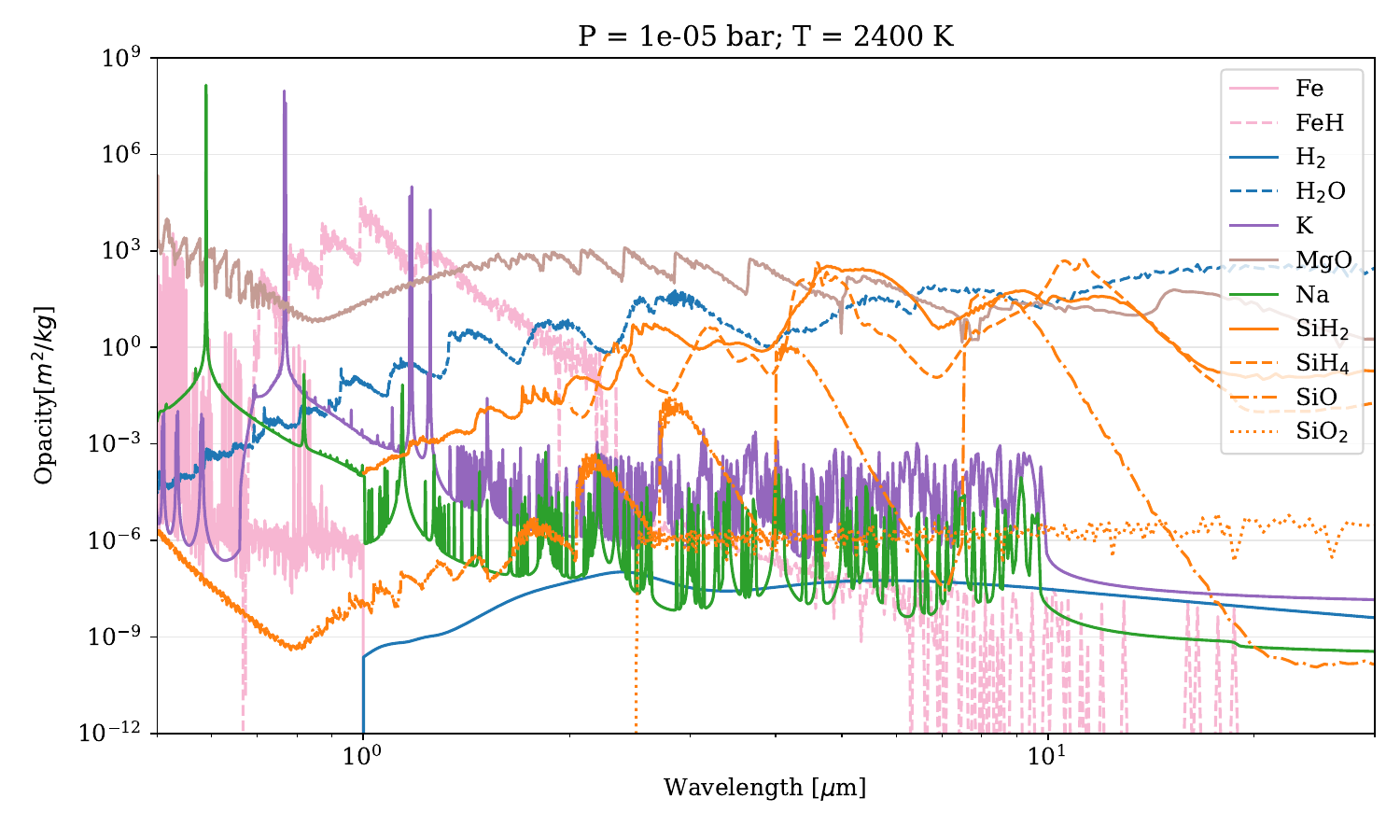}
  \caption{Opacity sources used in this study, at 0.1~bar (top), and $10^{-5}$~bar (bottom), both at 2400~K.
  }
  \label{Fig_opacities}%
\end{figure*}

Correlated-k opacities have mostly been computed through \textsc{ExoCross} \citep{yurchenko2018exocross} and SOCRATES \citep{edwards1996studies}, though
the SiOUVenIR opacity for SiO, and the opacities of \SiHF and \SiHt have been extracted from the \href{https://dace.unige.ch/opacityDatabase}{DACE database}\footnote{\url{https://dace.unige.ch/opacityDatabase}} and converted to correlated-k opacities via \texttt{Exo\_k} \citep{leconte2021}.
It should be noted that they have been generated in the context of gas giants, with an assumed $\HT$ and He dominated atmosphere.
This should have an effect on line broadening (see Fig.~1 from \cite{Amundsen2014} for example).
The opacities are shown at 2400~K at $10^{-5}$~bar and 1~bar in \fig{Fig_opacities}
for a resolution of 5000 spectral points between wavelengths $\lambda~=~0.2$ and 2000~$\mu$m.
All sources are listed in \tab{tab:opacity_sources}.
The resolution of the correlated-k opacities used is equally spaced in wavenumber and equal to 32 for the PT calculations (see \fig{Fig_PTs_phs} for example), while it is 5000 for the calculations of the spectra (see \fig{Fig_spectra_phs} for example).

\begin{table}[ht]\centering
\caption{Opacity sources}
    \begin{tabular}{c|c}
        Molecule & Line list \\\hline
        Na      & VALD \citep{Ryabchikova_2015} \\
        K       & VALD \citep{Ryabchikova_2015} \\
        Fe      & VALD \citep{Ryabchikova_2015} \\
        FeH     & MoLLIST \citep{Bernath_2020} \\
        \HT-\HT & HITRAN \citep{Gordon_2017} \\
        \HTO    & POKAZATEL \citep{Polyansky_2018} \\
        MgO     & LiTY \citep{Li_2019} \\
        SiO     & SiOUVenIR \citep{Yurchenko_2022} \\
        \SiOT   & OYT3 \citep{Owens_2020} \\
        \SiHF   & OY2T \citep{owens2017exomol} \\
        \SiHt   & CATS \citep{clark2020high}
    \end{tabular}
    \label{tab:opacity_sources}
\end{table}

\clearpage

\section{Optical depth}
\label{optical_depth}
\begin{figure}[ht]\centering
  \includegraphics[width=.5\textwidth]{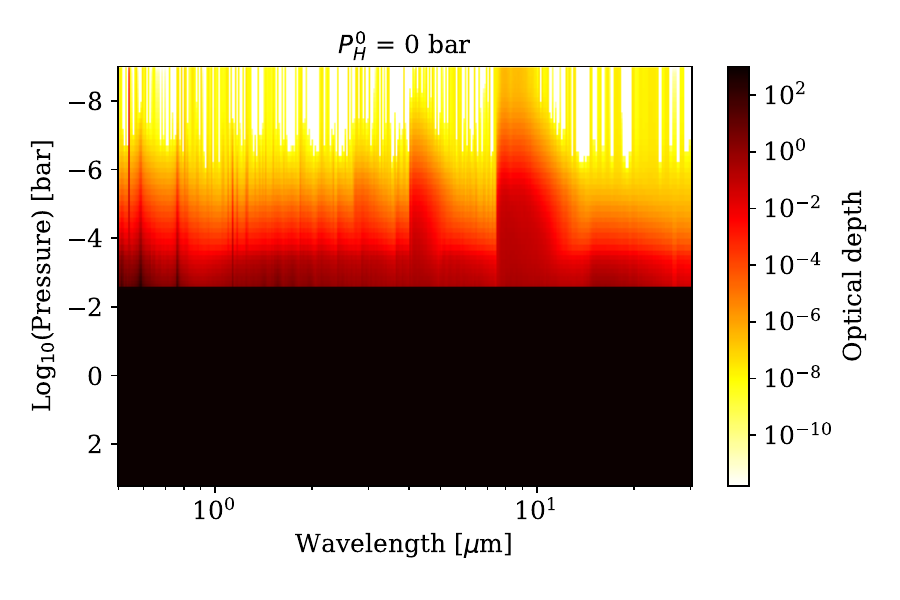}
  \includegraphics[width=.5\textwidth]{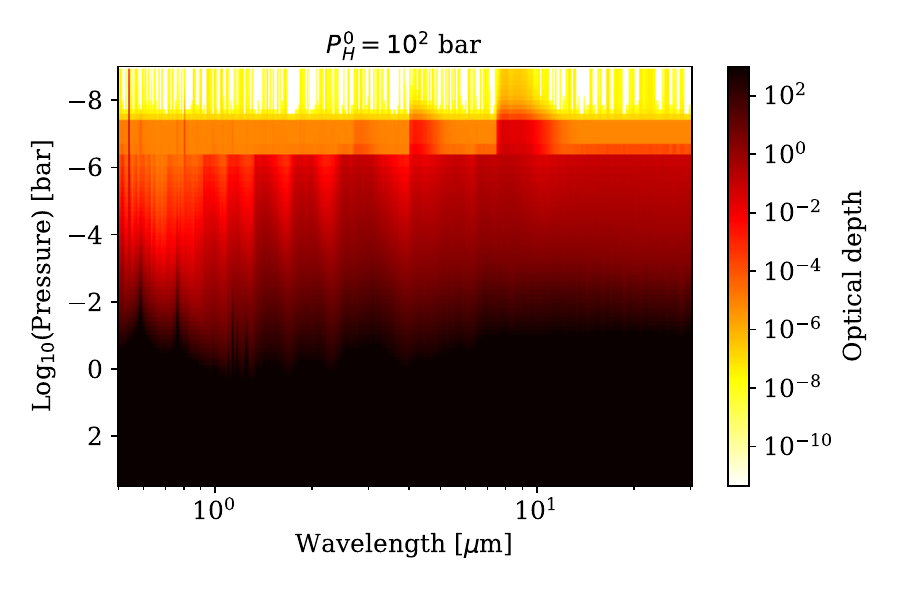}
  \caption{
  Optical depth for \Teq~=~2400~K, for a hydrogen-free case (top plot), and a hydrogen-rich case, i.e., \ph~=~$10^2$~bar (bottom plot).
  The color indicates the value of the optical depth, i.e., the cumulative opacity of the atmosphere. 
  Black is completely opaque while white is transparent.
}
  \label{Fig_optical_depth}%
\end{figure}

The optical depth can be written under the following form:
\begin{equation}
        \mathrm{d}\tau = -k(z)\mathrm{d}z,
        \label{eq:tau}
\end{equation}
where $k(z)$ is the sum of the opacities of all species weighted by their abundances.
The optical depth of the atmosphere can be shown per pressure level, such as shown in \fig{Fig_optical_depth}.
The figure shows two cases: one for a hydrogen-free case (top plot) and one for a hydrogen-rich case (with \ph~=$~10^2$~bar).
The two cases, with an equilibrium temperature \Teq of 2400~K, are quite distinct.
We can clearly see the interior boundary at a fixed pressure in the hydrogen-free case and the relatively thin layer of the atmosphere that is partially opaque.
We can see Na, K (in the optical) and SiO (in the infrared) spectral features, which we discuss more extensively in \sect{sec_spectral_features}.
The hydrogen-rich case has a more blurry boundary since the bottom of the atmosphere becomes opaque before the interior is reached.
The Na and K spectral features are still present, we can see the SiO spectral features at the top of the atmosphere, and \HTO bands between 1 and 5 $\mu$m.

\section{Spectral contribution of molecules}
\label{sec:contributions}

\begin{figure*}[ht]\centering
  \includegraphics[width=.8\textwidth]{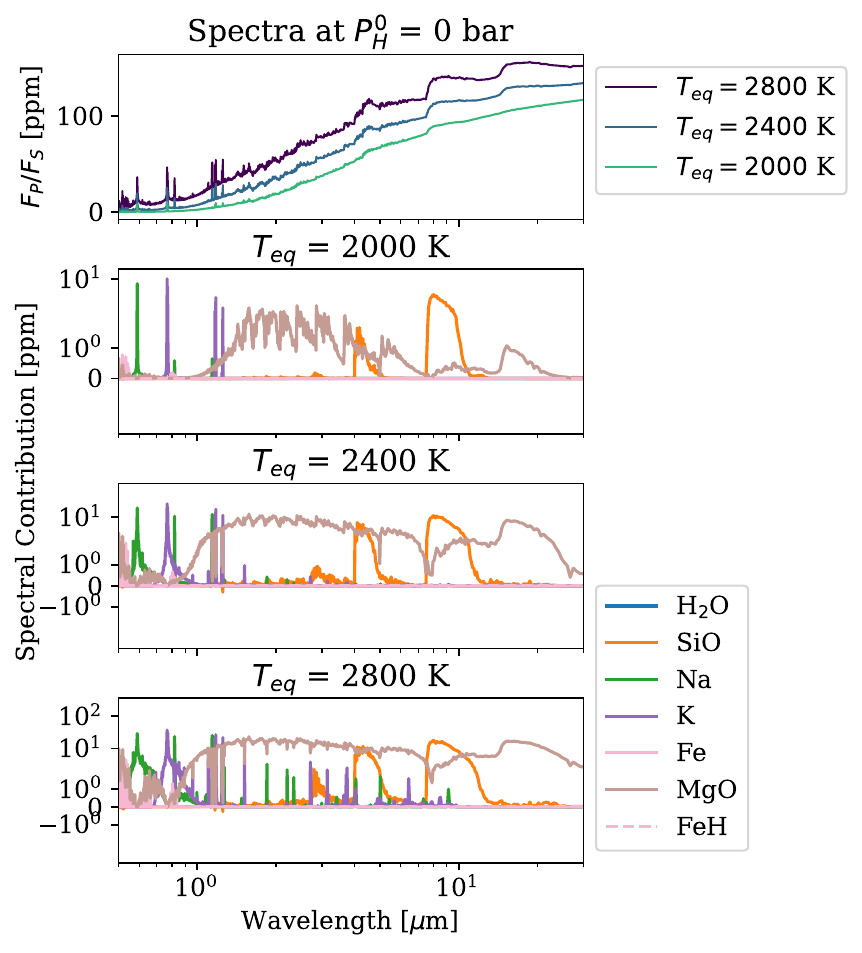}
  \caption{Spectral contributions of different molecules  without hydrogen and for different equilibrium temperatures \Teq. Positive contributions indicate emission features, while negative contributions indicate absorption features. The contribution of \HTO and FeH is null here.}
  \label{Fig_contribs_0}%
\end{figure*}

\fig{Fig_contribs_0},
to
\fig{Fig_contribs_1e+2} show the contributions of each molecule to the spectrum, for cases without hydrogen, with $P_H^0~=~10^{-2}$~bar and $P_H^0~=~10^{+2}$~bar, respectively, for the cases displayed in \app{sec:silicate_case} and \sect{sec:hydrogen_case}.
\begin{figure*}[ht]\centering
  \includegraphics[width=.8\textwidth]{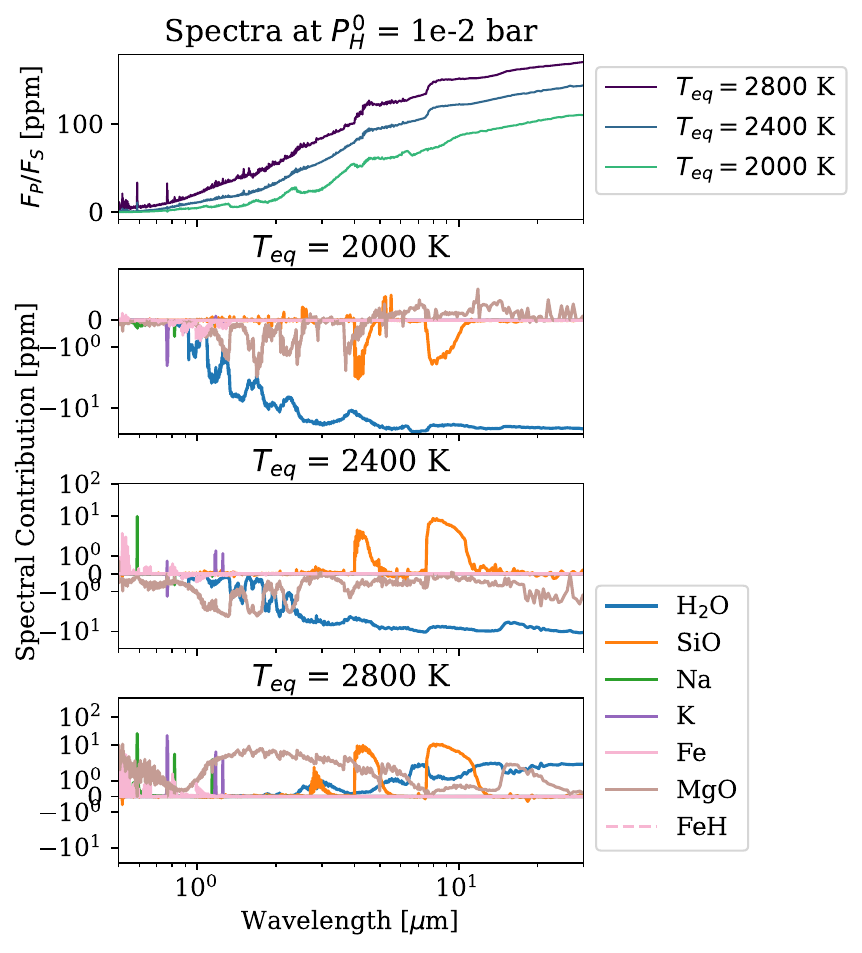}
  \caption{Spectral contributions of different molecules  at \ph=$10^{-2}$~bar and for different equilibrium temperatures \Teq.}
  \label{Fig_contribs_1e-2}%
\end{figure*}
\begin{figure*}[ht]\centering
  \includegraphics[width=.8\textwidth]{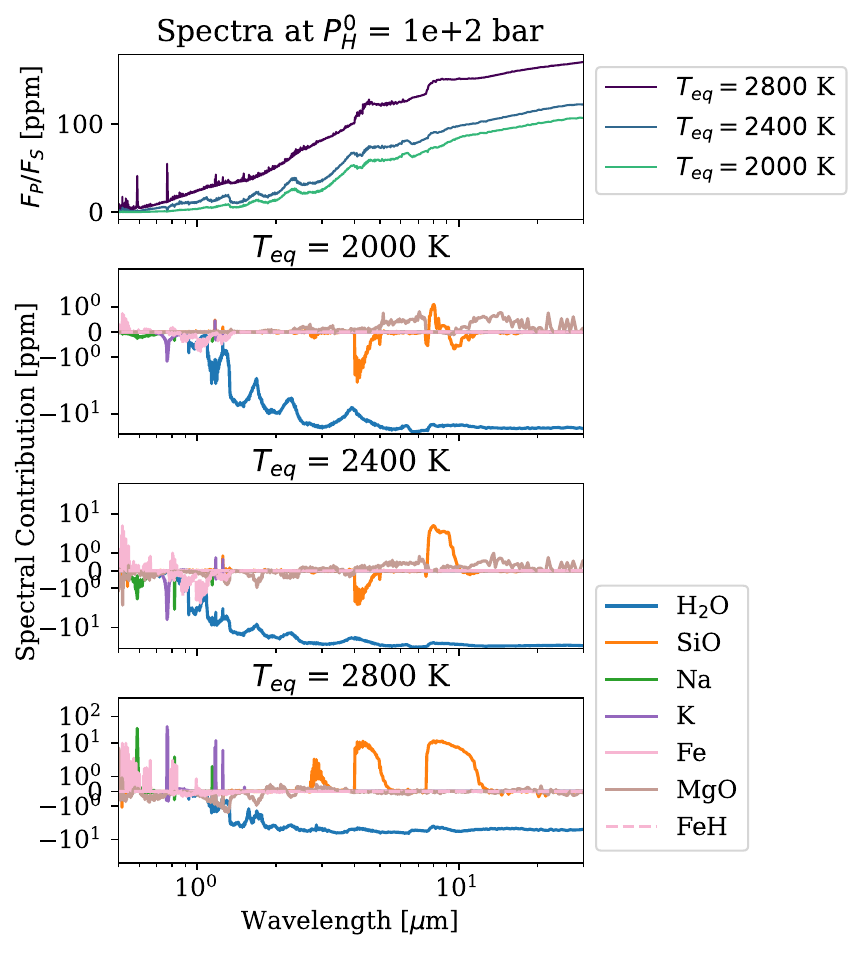}
  \caption{Spectral contributions of different molecules  at \ph=$10^{2}$~bar and for different equilibrium temperatures \Teq.}
  \label{Fig_contribs_1e+2}%
\end{figure*}

 \clearpage
 \clearpage

\section{Spectral contribution of \SiHt, \SiHF and \SiO at high hydrogen concentration}
\label{spectral_sih4}

The spectral contributions of \SiHt, \SiHF and \SiO are shown in \fig{Fig_contribs_1e+4} for \ph~=~$10^4$~bar.
\begin{figure}[ht]\centering
  \includegraphics[width=.5\textwidth]{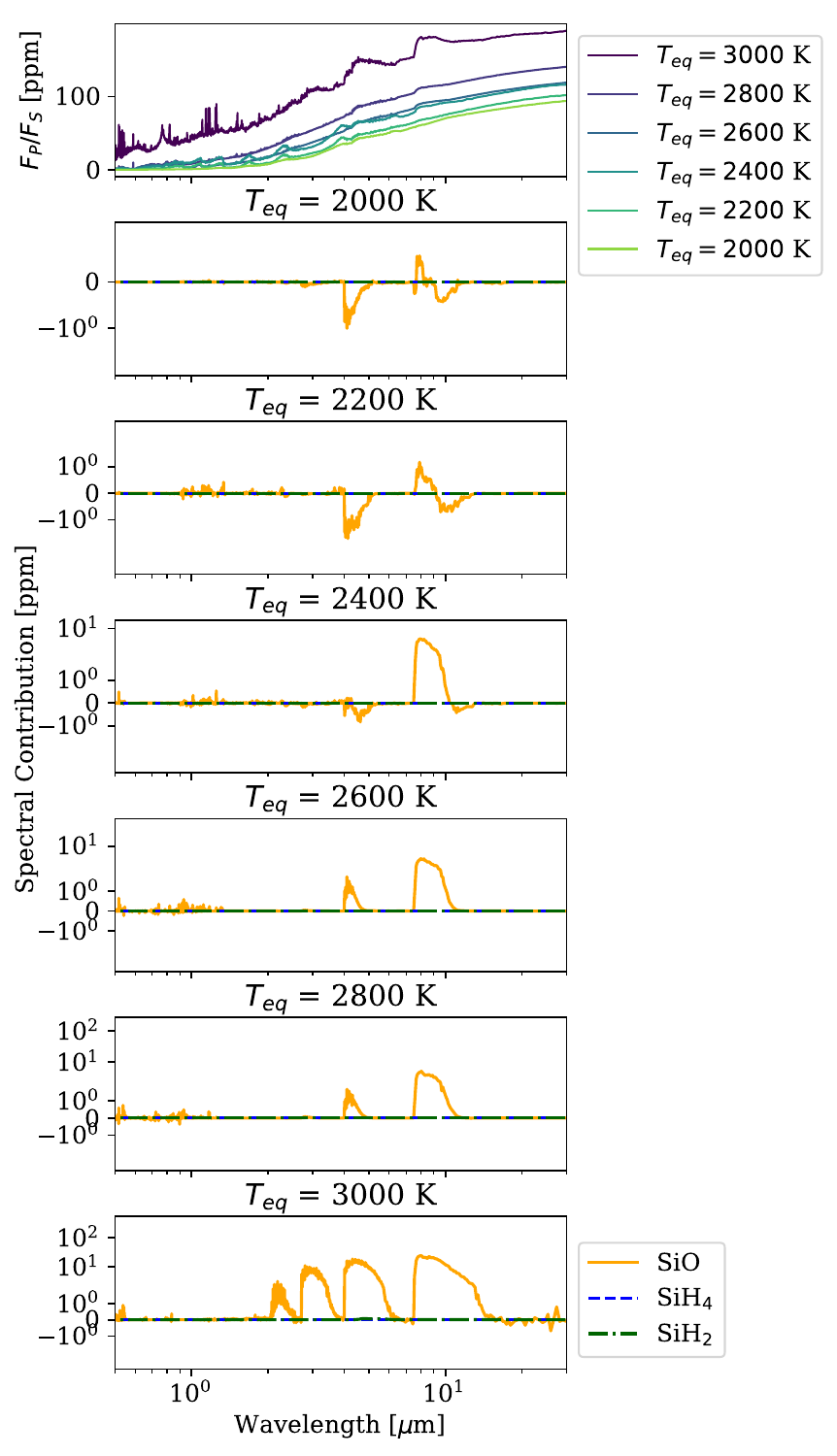}
  \caption{Spectral contributions of \SiO, \SiHF and \SiHt for \ph = $10^{4}$~bar of monoatomic hydrogen and for different equilibrium temperatures \Teq. \SiHF and \SiHt are negligible. \SiO is linked to very strong emission features at 2, 4, and 9 $\mu m$ at \Teq~=~3000~K.}
  \label{Fig_contribs_1e+4_6}%
\end{figure}
We can see that \SiHF and \SiHt have no significant spectral contribution, especially put in perspective with \SiO, which is coherent with their presence in the atmosphere being limited to the lower layers, near the surface.

The figure also shows how the spectral features of SiO are mixed between absorption features (mostly around 4 $\mu m$) for colder cases (low \Teq), and emission features (most prominent at $9 \mu m$) for hotter cases. 
The case at \Teq~=~3000~K is the only case corresponding to a thermal inversion, in which we can better see the emission peaks of SiO (at 1.5, 2, 4 and 9 $\mu m$).

\section{Spectral contribution when summing volatiles and vapor vs gas-liquid equilibrium}
\label{sec_contrib_sum}

\begin{figure}[ht]\centering
  \includegraphics[width=.5\textwidth]{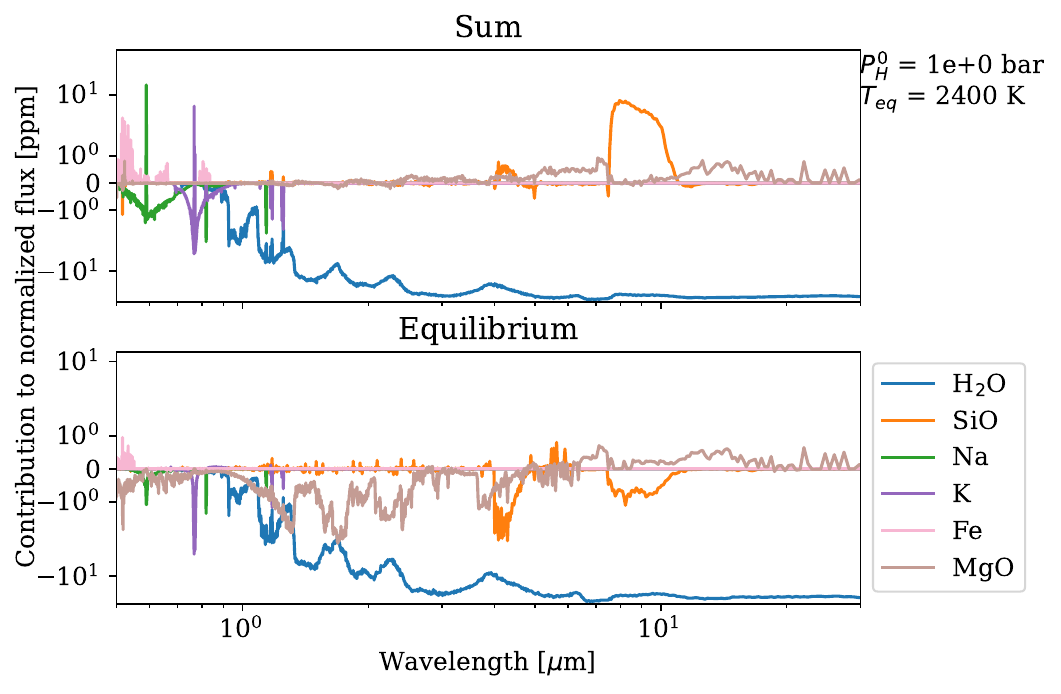}
  \caption{Spectral contributions for the \Sum and \Equilibrium methods, with \ph = $1$~bar and \Teq = 2400~K.}
  \label{Fig_contribs_sum}%
\end{figure}

We have compared in \sect{sec:sum_equilibrium} the \Sum and \Equilibrium methods, the first one computing the element abundances by simply summing the volatile abundances with the vapor abundances, and the second one computing a gas-liquid equilibrium.
\fig{Fig_contribs_sum} shows the spectral contributions of relevant species for the two methods.
We can see that \HTO is mainly the dominant absorbing species, but in the case of the \Sum method, a strong SiO feature is visible in emission, and also one Na and one K feature.
In the \Equilibrium case, these features disappear, due to the thermal inversion occurring at pressures that are too low (see \fig{Fig_pts_sum}).

\end{document}